\documentclass[12pt]{article}
\usepackage{amsmath,amssymb}
\usepackage{bm}
\usepackage{color}
\usepackage{graphicx}
\usepackage{cite}
\usepackage{mathtools}
\usepackage{breqn}

\usepackage{here}

\def\slash#1{\not\!\!#1}

\begin{document}

\title{
\begin{flushright}
\ \\*[-80pt]
\begin{minipage}{0.2\linewidth}
\normalsize
EPHOU-21-005\\*[50pt]
\end{minipage}
\end{flushright}
{\Large \bf
Majorana neutrino masses by D-brane instanton effects in 
magnetized orbifold models
\\*[20pt]}}

\author{
Kouki Hoshiya$^{a}$,
~Shota Kikuchi$^{a}$,
~Tatsuo Kobayashi$^{a}$, \\
 Kaito Nasu$^{a}$, 
~Hikaru Uchida$^{a}$, and 
~Shohei Uemura$^{b}$ 
\\*[20pt]
\centerline{
\begin{minipage}{\linewidth}
\begin{center}
{\it \normalsize
${}^{a}$Department of Physics, Hokkaido University, Sapporo 060-0810, Japan\\
${}^{b}$CORE of STEM, Nara Women's University, Nara 630-8506, Japan} \\*[5pt]
\end{center}
\end{minipage}}
\\*[50pt]}

\date{
\centerline{\small \bf Abstract}
\begin{minipage}{0.9\linewidth}
\medskip
\medskip
\small
We study Majorana neutrino masses induced by D-brane instanton effects in magnetized orbifold models.
We classify possible cases, where neutrino masses can be induced.
Three and four generations are favored in order to generate neutrino masses by D-brane instantons. Explicit mass matrices have specific features. Their diagonalizing matrices correspond to the bimaximal mixing matrix in the case with even magnetic fluxes, independently of the modulus value $\tau$. On the other hand, for odd magnetic fluxes, diagonalizing matrices correspond nearly to  the tri-bimaximal mixing matrix near $\tau =i$, while they become the bimaximal mixing matrix for larger ${\rm Im}\tau$. For even fluxes, neutrino masses are modular forms of the weight 1 on $T^2/\mathbb{Z}_2$, and they have symmetries such as $S_4'$ and ${\Delta'}(96)\times \mathbb{Z}_3$.
\end{minipage}
}

\begin{titlepage}
\maketitle
\thispagestyle{empty}
\end{titlepage}

\newpage


\section{Introduction}
\label{Intro}

Superstring theory is a promising candidate for the unified theory of 
all  the interactions including gravity, 
matter such as quarks and leptons, and the Higgs particle.
It predicts six dimensions  (6D) in addition to our four-dimensional (4D) spacetime.
The 6D space must be compact.
Geometrical aspects of the 6D compact space as well as gauge background fields 
determine phenomenological properties of particle physics such as generation numbers, 
flavor structure of quarks and leptons, and gauge and Yukawa coupling strengths.

Compactification with magnetic flux background is quite interesting in higher dimensional theory and 
superstring theory.
The magnetic flux background can lead to a 4D chiral theory even on the torus 
compactification \cite{Bachas:1995ik,Blumenhagen:2000wh,Angelantonj:2000hi,Blumenhagen:2000ea}, 
although one can not realize a 4D chiral theory on the torus without magnetic fluxes.
That is,  chiral zero-modes appear and its number depends on the size of magnetic flux.
Their wavefunctions are non-trivially quasi-localized.
Yukawa couplings as well as higher order couplings are written by 
overlap integrals of wavefunctions \cite{Cremades:2004wa,Abe:2009dr}.
They can be of ${\cal O}(1)$ or exponentially suppressed depending on 
distances of quasi-localizing positions among quarks, leptons and Higgs mode.

In addition to the simple torus compactification, 
the orbifold compactification with magnetic flux is quite 
interesting.
They can project out adjoint matter, i.e. open string moduli, and 
lead to the numbers of chiral zero-modes different from the torus compactifictaion \cite{Abe:2008fi,Abe:2013bca,Abe:2014noa} 
\footnote{See also for the numbers of zero-modes Refs.~\cite{Kobayashi:2017dyu,Sakamoto:2020pev}.}.
One can construct various models by properly choosing orbifold parities 
and Sherk-Schwartz (SS) phases as well as discrete Wilson lines \footnote{See for shifted orbifold models  \cite{Fujimoto:2013xha}.}.
The three-generation models in magnetized orbifold models 
have been classified \cite{Abe:2008sx,Abe:2015yva,Hoshiya:2020hki}.
Furthermore, realistic quark  masses and mixing angles as well as the CP phase were studied 
\cite{Abe:2012fj,Abe:2014vza,Fujimoto:2016zjs,Kobayashi:2016qag}.
Also realization of charged lepton masses was studied.
Recently, their flavor structure was studied from the viewpoint of 
modular symmetry \cite{Kobayashi:2018rad,Kobayashi:2018bff,Kariyazono:2019ehj,Ohki:2020bpo,Kikuchi:2020frp,Kikuchi:2020nxn,Kikuchi:2021ogn,Almumin:2021fbk}.

Small neutrino masses can be realized by the see-saw mechanism as well as the Weinberg operators.
Here, we concentrate on the see-saw mechanism, 
where right-handed neutrinos and their Majorana masses are introduced.
Such Majorana masses can be induced by non-perturbative effects, 
i.e. D-brane instanton effects \cite{Blumenhagen:2006xt,Ibanez:2006da}, although there is no mass scale perturbatively below 
the compactification scale in superstring theory and super Yang-Mills theory, 
which is low-energy effective field theory of superstring theory.
(See also for explicit models Refs.~\cite{Ibanez:2007rs,Antusch:2007jd,Kobayashi:2015siy}.)
It is very important to study explicitly which patterns of Majorana masses are 
induced by D-brane instanton effects.
Thus, our purpose in this paper is to study systematically patterns of Majorana mass matrices 
induced by D-brane effects in magnatized orbifold models, 
in particular $T^2/\mathbb{Z}_2$ orbifold models.

When a D-brane instanton appears, there appear new zero-modes, $\beta_i$ and $\gamma_j$, which correspond to 
open strings between the D-brane instanton and the D-branes for the right-handed neutrinos.
We integrate the new zero-modes so as to obtain non-perturbative correction terms.
The Majorana mass terms can be induced only when we have a certain number of the new zero-modes, $\beta_i$ and $\gamma_j$.
In magnetized orbifold models, the number of zero-modes is determined by 
the size of magnetic fluxes, $\mathbb{Z}_2$-parity and SS phases.
In this paper, we study systematically D-brane configurations leading to Majorana neutrino masses 
and compute the patterns of neutrino mass matrices.

This paper is organized as follows.
In section \ref{sec:Revew}, we give a brief review on 
the number of zero-modes, their wavefunctions, and 3-point couplings.
In section \ref{sec:D-instanton}, we systematically study 
D-brane configurations, where 
Majorana neutrino masses can be induced by 
D-brane instanton effects.
In section \ref{sec:Neutrino-mass}, we compute explicitly 
the patterns of Majorana neutrino mass matrices.
Section \ref{sec:conclusion} is conclusion.
In Appendix \ref{app:moduar}, we review briefly the modular symmetry 
in magnetized torus and orbifold models.
In Appdendix \ref{app:Z4}, we review on wave functions on the $T^2/\mathbb{Z}_4$ orbifold.


\section{Orbifold compactification with magnetic fluxes}
\label{sec:Revew}

Here we review torus and orbifold compactfications with magnetic fluxes~\cite{Cremades:2004wa,Abe:2008fi,Abe:2013bca,Abe:2014noa}.
We study magnetic flux compactification within the framework of 
higher dimensional super Yang-Mills theory, 
which is low-energy effective field theory of superstring theory.
We concentrate on extra two dimensions, $T^2$ and $T^2/\mathbb{Z}_2$.
Our models can correspond to the D5-brane system as well as D7 and D9-brane systemes, 
assuming that other extra dimensions are irrelevant to mass ratios and mixing angles of 
Majorana neutrino mass matrix, but relevant only to the overall factor of 
the mass matrix.

\subsection{Torus compactification with magnetic fluxes}

In this subsection, we review zero-mode wavefunctions on a two-dimensional torus $T^2$ with $U(N)$ magnetic flux.
First, $T^2$ is obtained by the identification, $z \sim z+1 \sim z+\tau$, where we use the complex coordinate $z \equiv x + \tau y$ and $\tau$ ($\tau \in \mathbb{C}$, ${\rm Im}\tau > 0$) is the complex structure modulus of $T^2$.
Let us consider the following $U(N)$ magnetic flux on the $T^2$,
\begin{align}
F = \frac{\pi i}{{\rm Im}\tau} (dz \wedge d\bar{z})
\begin{pmatrix}
M_1 \mathbb{I}_{N_1 \times N_1} & & \\
 & \ddots & \\
 & & M_n \mathbb{I}_{N_n \times N_n}
\end{pmatrix},
\label{F}
\end{align}
where $\mathbb{I}_{N_a \times N_a}$ $(a=1,...,n)$ denotes the $(N_a \times N_a)$ unit matrix with $\sum_{a=1}^{n}N_a=N$ and $M_a$ $(a=1,...,n)$ must be integer.
This magnetic flux is induced by the background gauge field,
\begin{align}
A(z) = \frac{\pi}{{\rm Im}\tau} {\rm Im}\left( \bar{z}dz \right)
\begin{pmatrix}
M_1 \mathbb{I}_{N_1 \times N_1} & & \\
 & \ddots & \\
 & & M_n \mathbb{I}_{N_n \times N_n}
\end{pmatrix},
\label{A}
\end{align}
where we can always set Wilson lines vanishing by introducing SS phases instead, as shown in Ref.~\cite{Abe:2013bca}.
From this gauge background, $U(N)$ gauge symmetry is broken to $\Pi_{a=1}^{n}U(N_a)$.

Here, we consider two-dimensional spinor on $T^2$ under breaking $U(N) \rightarrow U(N_1) \times U(N_2)$ by the above background magnetic flux, that is,
\begin{align}
\psi(z) =
\begin{pmatrix}
\psi_+(z) \\ \psi_-(z)
\end{pmatrix}, \quad
\psi_{\pm}(z) =
\begin{pmatrix}
\psi_{\pm}^{(11)}(z) & \psi_{\pm}^{(12)}(z) \\
\psi_{\pm}^{(21)}(z) & \psi_{\pm}^{(22)}(z)
\end{pmatrix},
\end{align}
where $\psi_{\pm}^{(11)}$ and $\psi_{\pm}^{(22)}$ correspond to $U(N_1)$ and $U(N_2)$ gauginos, respectively and $\psi_{\pm}^{(12)}$ and $\psi_{\pm}^{(21)}$ correspond to $(N_1, \bar{N}_2)$ and $(\bar{N}_1, N_2)$ representations under $U(N_1) \times U(N_2)$.
The above wavefunctions satisfy the following boundary conditions,
\begin{align}
&\psi_{\pm}^{(ab)}(z+1) = e^{2\pi i \alpha_1^{(ab)}}e^{\pi iM_{ab} \frac{{\rm Im}z}{{\rm Im}\tau}} \psi_{\pm}^{(ab)}(z), \label{psiz1} \\
&\psi_{\pm}^{(ab)}(z+\tau) = e^{2\pi i \alpha_{\tau}^{(ab)}}e^{\pi iM_{ab} \frac{{\rm Im}\bar{\tau}z}{{\rm Im}\tau}} \psi_{\pm}^{(ab)}(z), \label{psiztau}
\end{align}
where $M_{ab} \equiv M_a -M_b$ ($a,b=1,2$) and $\alpha_i^{(ab)} \equiv \alpha_i^{a} - \alpha_i^{b}$ ($i=1,\tau$, $a,b=1,2$) denote the SS phases.
When we solve the zero-mode Dirac equation,
\begin{align}
  i\slash{D}\psi(z) = 0,
\end{align}
under the above boundary conditions with $M_{12}>0$, only $\psi_+^{(12)}$ (as well as the anti-particle $\psi_-^{(21)}$) has $M_{12}$-number of degenerate zero-modes.
Then, a chiral theory is realized.
The explicit form of the wavefunction with the magnetic flux $M_{12}$ and the SS phases ($\alpha_1^{(12)}, \alpha_{\tau}^{(12)}$) is 
written by 
\begin{align}
\psi_{T^2}^{(j+\alpha_1^{(12)},\alpha_{\tau}^{(12)}),|M_{12}|}(z)
&= \left(\frac{|M_{12}|}{{\cal A}^2}\right)^{1/4} e^{2\pi i\frac{(j+\alpha_1^{(12)})\alpha_{\tau}^{(12)}}{|M_{12}|}} e^{\pi i|M_{12}|z \frac{{\rm Im}z}{{\rm Im}\tau}}
\vartheta
\begin{bmatrix}
\frac{j+\alpha_1^{(12)}}{|M_{12}|}\\
-\alpha_{\tau}^{(12)}
\end{bmatrix}
(|M_{12}|z, |M_{12}|\tau),
\label{psiT2}
\end{align}
where $j \in \mathbb{Z}_{|M_{12}|}$, ${\cal A}$ denotes the area of $T^2$, and $\vartheta$ denotes the Jacobi theta function defined as
\begin{align}
\vartheta
\begin{bmatrix}
a\\
b
\end{bmatrix}
(\nu, \tau)
=
\sum_{l\in \mathbb{Z}}
e^{\pi i (a+l)^2\tau}
e^{2\pi i (a+l)(\nu+b)}.
\end{align}
The wavefunction satisfies the following normalization condition,
\begin{align}
  \int_{T^2} dzd\bar{z} \left(\psi^{(j+\alpha_1^{(12)},\alpha_{\tau}^{(12)}),|M_{12}|}_{T^2}(z)\right)^* \psi^{(k+\alpha_1^{(12)},\alpha_{\tau}^{(12)}),|M_{12}|}_{T^2}(z) 
  = (2{\rm Im}\tau)^{-1/2} \delta_{j,k}.
\end{align}
We note that the definition of the wavefunction in Eq.~(\ref{psiT2}) is different from one in Ref.~\cite{Abe:2013bca} by the phase factor, $e^{2\pi i (j+\alpha_1^{(12)})\alpha_{\tau}^{(12)}/|M_{12}|}$, which does not affect the boundary conditions or the equation of motion.
Similarly, we can consider the breaking case such as $U(N) \rightarrow \Pi_{a=1}^{n}U(N_a)$.

The three-point coupling, $d^{ijk}$, of the wavefunctions with the magnetic flux $M_a$ and the SS phases $(\alpha_1^{(a)}, \alpha_{\tau}^{(a)})$ $(a=i,j,k)$ can be calculated as
\begin{align}
d^{ijk}
&= \int_{T^2} dz d\bar{z} \psi_{T^2}^{(i+\alpha_1^{(i)},\alpha_{\tau}^{(i)}),|M_i|}(z) \cdot \psi_{T^2}^{(j+\alpha_1^{(j)},\alpha_{\tau}^{(j)}),|M_j|}(z) \cdot \left(\psi_{T^2}^{(k+\alpha_1^{(k)},\alpha_{\tau}^{(k)}),|M_k|}(z)\right)^* \notag \\
&= c_{(M_i - M_j - M_k)}\exp \left\{ 2\pi i \left( \frac{(i+\alpha_1^{(i)})\alpha_{\tau}^{(i)}}{M_i} + \frac{(j+\alpha_1^{(j)})\alpha_{\tau}^{(j)}}{M_j} - \frac{(k+\alpha_1^{(k)})\alpha_{\tau}^{(k)}}{M_k} \right) \right\} \notag \\
& \times \sum_{m=0}^{|M_k|-1} \vartheta
  \begin{bmatrix}
    \frac{|M_j|(i+\alpha_1^{(i)})-|M_i|(j+\alpha_1^{(j)})+|M_iM_j|m}{|M_iM_jM_k|} \\ 0
  \end{bmatrix}
  (M_i\alpha_{\tau}^{(j)}-M_j\alpha_{\tau}^{(i)}, |M_iM_jM_k|\tau) \notag \\ 
 \label{eq: yukawa}
&\times \delta_{(i+\alpha_1^{(i)})+(j+\alpha_1^{(j)})-(k+\alpha_1^{(k)}),|M_k|\ell-|M_i|m},
\end{align}
provided $M_i+M_j=M_k$, $\alpha_1^{(i)}+\alpha_1^{(j)}=\alpha_1^{(k)}$, and $\alpha_{\tau}^{(i)}+\alpha_{\tau}^{(j)}=\alpha_{\tau}^{(k)}$ are satisfied, where $\ell \in \mathbb{Z}$. Here, the coefficient $c_{(M_i - M_j - M_k)}$ is defined as
\begin{equation}
\label{eq: overall constant}
c_{(M_i - M_j - M_k)} = (2{\rm Im}\tau)^{-1/2} {\cal A}^{-1/2} \left|\frac{M_iM_j}{M_k}\right|^{1/4}.
\end{equation}
To derive Eq. (\ref {eq: yukawa}), we use the property
\begin{align}
  \vartheta \begin{bmatrix} \frac{r}{N_1} \\ 0 \end{bmatrix} (\nu_1,N_1\tau) \times \vartheta \begin{bmatrix} \frac{s}{N_2} \\ 0 \end{bmatrix} (\nu_2,N_2\tau) 
  &= \sum_{m\in\mathbb{Z}_{N_1+N_2}} \vartheta \begin{bmatrix} \frac{r+s+N_1m}{N_1+N_2} \\ 0 \end{bmatrix} (\nu_1+\nu_2,(N_1+N_2)\tau) \notag \\
  \times \vartheta &\begin{bmatrix} \frac{N_2r-N_1s+N_1N_2m}{N_1N_2(N_1+N_2)} \\ 0 \end{bmatrix} (\nu_1N_2-\nu_2N_1,N_1N_2(N_1+N_2)\tau).
\end{align}

\subsection{Orbifold compactification}
\label{sec:orbifold}
In this subsection, we review zero-mode wavefunctions on the $T^2/\mathbb{Z}_2$ twisted orbifold with magnetic flux in Eq.~(\ref{F}).
The $T^2/\mathbb{Z}_2$ twisted orbifold is obtained further by identifying $\mathbb{Z}_2$ twisted point $-z$ with $z$, i.e.~$z \sim -z$.
Then, the wavefunctions on magnetized $T^2/\mathbb{Z}_2$ twisted orbifold with the magnetic flux $M$ and the SS phases $(\alpha_1, \alpha_{\tau})$ satisfy the following boundary condition,
\begin{align}
\psi^{(j+\alpha_1,\alpha_{\tau}),|M|}_{T^2/\mathbb{Z}_2^{m}}(-z)  = (-1)^m \psi^{(j+\alpha_1,\alpha_{\tau}),|M|}_{T^2/\mathbb{Z}_2^{m}}(z), \quad m \in \mathbb{Z}_2,
\end{align}
in addition to Eqs.~(\ref{psiz1}) and (\ref{psiztau}).
These boundary conditions are simultaneously satisfied only in the case of $\mathbb{Z}_2$ SS phases,
\begin{align}
(\alpha_1, \alpha_{\tau}) = (0,0), (1/2,0), (0,1/2), (1/2,1/2).
\end{align}
Hence, the wavefunctions on magnetized $T^2/\mathbb{Z}_2$ twisted orbifold can be expressed by ones on magnetized $T^2$ as
\begin{align}
\psi^{(j+\alpha_1,\alpha_{\tau}),|M|}_{T^2/\mathbb{Z}_2^{m}}(z)
&= {\cal N} \left(\psi_{T^2}^{(j+\alpha_1,\alpha_{\tau}),|M|}(z) + (-1)^m \psi_{T^2}^{(j+\alpha_1,\alpha_{\tau}),|M|}(-z)\right) \notag \\
&= {\cal N} \left(\psi_{T^2}^{(j+\alpha_1,\alpha_{\tau}),|M|}(z) + (-1)^{m-2\alpha_{\tau}} \psi_{T^2}^{\left(|M|-(j+\alpha_1),\alpha_{\tau} \right),|M|}(z)\right), \label{psiZ2}
\end{align}
where
\begin{align}
 {\cal N} = \left\{
  \begin{array}{l}
    1/2 \quad (j+\alpha_1=0,|M|/2) \\
    1/\sqrt{2} \quad ({\rm otherwise})
  \end{array}
  \right..
\end{align}
Note that only when $j+\alpha_1=0$, the factor $(-1)^{2\alpha_1}$ is replaced by $1$.
Tables \ref{00}, \ref{10}, \ref{01}, and \ref{11} show the numbers of the zero-modes.
\begin{table}[H]
\centering
\begin{tabular}{|c|c|c|c|c|c|c|c|c|c|c|c|c|} \hline
$|M|$ & 1 & 2 & 3 & 4 & 5 & 6 & 7 & 8 & 9 & 10 & 11 & 12 \\ \hline
$\mathbb{Z}_2$-even & 1 & 2 & 2 & 3 & 3 & 4 & 4 & 5 & 5 & 6 & 6 & 7 \\ \hline
$\mathbb{Z}_2$-odd & 0 & 0 & 1 & 1 & 2 & 2 & 3 & 3 & 4 & 4 & 5 & 5 \\ \hline
\end{tabular}
\caption{The number of zero-modes with the SS phase $(\alpha_1, \alpha_{\tau}) = (0,0)$.}
\label{00}
\centering
\begin{tabular}{|c|c|c|c|c|c|c|c|c|c|c|c|c|} \hline
$|M|$ & 1 & 2 & 3 & 4 & 5 & 6 & 7 & 8 & 9 & 10 & 11 & 12 \\ \hline
$\mathbb{Z}_2$-even & 1 & 1 & 2 & 2 & 3 & 3 & 4 & 4 & 5 & 5 & 6 & 6 \\ \hline
$\mathbb{Z}_2$-odd & 0 & 1 & 1 & 2 & 2 & 3 & 3 & 4 & 4 & 5 & 5 & 6 \\ \hline
\end{tabular}
\caption{The number of zero-modes with the SS phase $(\alpha_1, \alpha_{\tau}) = (1/2,0)$.}
\label{10}
\centering
\begin{tabular}{|c|c|c|c|c|c|c|c|c|c|c|c|c|} \hline
$|M|$ & 1 & 2 & 3 & 4 & 5 & 6 & 7 & 8 & 9 & 10 & 11 & 12 \\ \hline
$\mathbb{Z}_2$-even & 1 & 1 & 2 & 2 & 3 & 3 & 4 & 4 & 5 & 5 & 6 & 6 \\ \hline
$\mathbb{Z}_2$-odd & 0 & 1 & 1 & 2 & 2 & 3 & 3 & 4 & 4 & 5 & 5 & 6 \\ \hline
\end{tabular}
\caption{The number of zero-modes with the SS phase $(\alpha_1, \alpha_{\tau}) = (0,1/2)$.}
\label{01}
\centering
\begin{tabular}{|c|c|c|c|c|c|c|c|c|c|c|c|c|} \hline
$|M|$ & 1 & 2 & 3 & 4 & 5 & 6 & 7 & 8 & 9 & 10 & 11 & 12 \\ \hline
$\mathbb{Z}_2$-even & 0 & 1 & 1 & 2 & 2 & 3 & 3 & 4 & 4 & 5 & 5 & 6 \\ \hline
$\mathbb{Z}_2$-odd & 1 & 1 & 2 & 2 & 3 & 3 & 4 & 4 & 5 & 5 & 6 & 6 \\ \hline
\end{tabular}
\caption{The number of zero-modes with the SS phase $(\alpha_1, \alpha_{\tau}) = (1/2,1/2)$.}
\label{11}
\end{table}
The normalization of the wavefunction and the three-point coupling of them are similarly obtained by replacing the wavefunctions on $T^2$ with ones on the $T^2/\mathbb{Z}_2$ twisted orbifold.
Note that they are overall $\mathbb{Z}_2$ invariant.

\section{Majorana neutrino masses by D-brane instanton effects}
\label{sec:D-instanton}

Here, we study Majorana neutrino masses induced by D-brane instanton effects 
in magnetized orbifold models.

\subsection{D-brane instanton effects}

Here, we give a brief review on Majorana neutrino mass terms induced by D-brane instanton effects 
\cite{Blumenhagen:2006xt,Ibanez:2006da}.

We consider two stacks of D-branes D$_{N1}$ and D$_{N2}$ with different magnetic fluxes.
Zero-modes of open strings between these D-branes correspond to the neutrinos, $N_a$.
We denote the difference of their magnetic fluxes $M_N$.
The neutrino generation number is determined by $M_N$ and boundary condition such as SS phases and 
$\mathbb{Z}_2$ parity.
We assume the D-brane instanton D$_{inst}$ with magnetic flux, which has the zero-modes 
$\beta_i$ ($\gamma_j$) between D$_{N1}$ (D$_{N2}$) and D$_{inst}$.
The numbers of zero-modes $\beta_i$ and $\gamma_j$ are determined by 
magnetic fluxes in zero-mode equations, and  SS phases and $\mathbb{Z}_2$ parities of boundary conditions.

The Majorana mass terms of $N_a$ due to D-brane instanton effects can be written by\cite{Blumenhagen:2006xt,Ibanez:2006da}
\begin{eqnarray}
e^{-S_{cl}(D_{inst},M_{inst})}\int d^2\beta d^2\gamma e^{-\sum_{ija} d^{ij}_a\beta_i \gamma_j N_a},
\end{eqnarray}
where $\beta_i$ and $\gamma_j$ are  grassmannian, and $d^{ij}_a$ denotes  the 3-point coupling among 
$\beta_i$, $\gamma_j$, $N_a$.
Here,  $S_{cl}(D_{inst},M_{inst})$ denotes 
the classical action of the D-brane instanton written by the Dirac-Born-Infeld action, which depends on the 
D-brane instanton volume in the compact space and the magnetic flux.
Mass terms can be induced only if each of $\beta_i$ and $\gamma_j$ 
has two zero-modes, i.e. $\beta_1$, $\beta_2$, $\gamma_1$ and $\gamma_2$.
By the grassmannian integral, we find
 \begin{eqnarray}
\int d^2\beta d^2\gamma e^{-\sum_{ija} \beta_i \gamma_j N_a}=N_aN_{b}(\varepsilon_{ij}\varepsilon_{k\ell} d^{ik}_{a}d^{j\ell}_b).
\end{eqnarray}
Thus the mass matrix $M_{ab}$ is obtained by 
\begin{eqnarray}
M_{ab}= e^{-S_{cl}(D_{inst},M_{inst})}m_{ab}, \qquad m_{ab}=(\varepsilon_{ij}\varepsilon_{k\ell} d^{ik}_{a}d^{j\ell}_b).
\end{eqnarray}

\subsection{Classification for models with Majorana mass terms}

The number of zero-modes is required to equal to two for both $\beta_i$ and $\gamma_j$ 
in order to induce Majorana neutrino mass terms by D-brane instanton effects 
as reviewed in the previous section.
We consider the six-dimensional compact space, which is a product of $T^2/\mathbb{Z}_2$ and four-dimensional 
compact space.
Here, we focus on the two-dimensional $T^2/\mathbb{Z}_2$ orbifold, and 
we study the models, which has two zero-modes, $\beta_i$ and $\gamma_j$.
When the instanton brane D$_{ints}$ wraps on other four-dimensional compact space, 
each of $\beta_i$ and $\gamma_j$ must have only one zero-mode on other four-dimensional compact space 
such that the total zero-mode number is equal to two.
Note that the total zero-mode number is a product of zero-mode numbers on $T^2/\mathbb{Z}_2$ and 
other four-dimensional compact space.
Thus, the flavor structure, i.e. mass ratios and mixing angles, is determined by 
configuration on $T^2/\mathbb{Z}_2$, although other four-dimensional compact space 
contributes only on an overall factor of mass matrix.

Two zero-modes can be realized by D-brane instantons with proper magnetic fluxes 
and SS phases as reviewed in section \ref{sec:orbifold}.
These are shown in Table \ref{tab:zero-mode-alpha}.
In addition, the neutrinos must have non-vanishing 3-point couplings  $d^{ij}_a$ with 
the zero-modes, $\beta_i$ and $\gamma_j$.
That leads the following conditions:
\begin{equation}
M_\beta \pm M_\gamma = \pm M_N,  \qquad (\alpha_1,\alpha_\tau)_\beta + (\alpha_1,\alpha_\tau)_\gamma = 
(\alpha_1,\alpha_\tau)_N,
\end{equation}
where $M_\beta$, $M_\gamma$, $M_N$ are magnetic fluxes in zero-mode equations of 
$\beta_i$, $\gamma_i$, neutrinos $N_a$, and 
$ (\alpha_1,\alpha_\tau)_\beta$, $(\alpha_1,\alpha_\tau)_\gamma$, 
$(\alpha_1,\alpha_\tau)_N$ are SS phases in boundary conditions of 
$\beta_i$, $\gamma_j$, neutrinos $N_a$, respectively.
Note that the SS phases $(\alpha_1,\alpha_\tau)$ are defined modulo integer.
Furthermore, 3-point couplings $d^{ij}_a$ are allowed only if the product of $\mathbb{Z}_2$ parities 
of  $\beta_i$, $\gamma_j$, and neutrinos $N_a$, is $\mathbb{Z}_2$-even.
Thus, when we fix magnetic fluxes, SS phases and $\mathbb{Z}_2$ parities for $\beta_j$ and $\gamma_j$, 
those are determined for neutrinos.
Then, we can find the generation number of neutrinos, which can gain mass terms 
through the D-brane instanton effects.
Such generation numbers are shown in Table \ref{tab:n-generation} for $M_N=M_{\beta}+M_{\gamma}$.
When $M_N=|M_{\beta}-M_{\gamma}|$, the neutrino generation number is 0 or 1.

\begin{table}[h]
\begin{center}
\begin{tabular}{|c|c|} \hline
$\beta_i, \gamma_j $ & ($\alpha_1,\alpha_\tau:M_{\beta,\gamma}$,$\mathbb{Z}_2$ parity)\\ \hline
i &(0,0:2,E) \\ \hline
ii& (0,0:3,E)  \\ \hline
iii& (0,0:5,O) \\ \hline
iv& (0,0:6,O) \\ \hline
v& ({1/2,0}:3,E)\\ \hline
vi & ({0,1/2}:3,E)\\ \hline
vii& ({1/2,0}:4,E)\\ \hline
viii & ({0,1/2}:4,E)\\ \hline
ix & ({1/2,0}:4,O)\\ \hline
x& (0,{1/2}:4,O) \\ \hline
xi& ({1/2,0}:5,O)\\ \hline
xii& (0,{1/2}:5,O)\\ \hline
xiii& (1/2,1/2:4,E)\\ \hline
xiv& (1/2,1/2:5,E) \\ \hline
xv& (1/2,1/2:3,O) \\ \hline
xvi& (1/2,1/2:4,O)\\ \hline
\end{tabular}
\end{center}
\caption{Configurations leading to two zero-modes for  $\beta_i$ and $\gamma_j$.}
\label{tab:zero-mode-alpha}
\end{table}

\begin{table}[h]
\begin{center}
\begin{tabular}{|c|c|c|c|c|c|c|c|c|c|c|c|c|c|c|c|c|} \hline
$\beta_i \backslash \gamma_j $ & i & ii &iii & iv & v & vi & vii& viii& ix& x & xi & xii & xiii & xiv & xv & xvi \\ \hline
i& 3& 3& 3& 3& 3& 3& 3& 3& 3& 3& 3& 3& 3& 3& 3&3 \\ \hline
ii & 3& 4&3 &4 &3 &3 &4 &4 &3 &3 &4 & 4&3 &4 &3 &4 \\ \hline
iii & 3&3 & 6& 6& 4&4 &4 &4 &5 &5 &5 &5 &5 &5 &4 &4 \\ \hline
iv & 3& 4& 6&7 &4 &4 &5 &5 &5 &5 &6 &6 &5 &6 &4 &5 \\ \hline
v& 3&3 &4 & 4& 4& 3& 4& 3& 3& 4& 3& 4& 4& 4&3 & 3\\ \hline
vi& 3& 3& 4& 4& 3& 4& 3& 4& 4& 3&4 &3 & 4& 4& 3& 3\\ \hline
vii& 3& 4& 4& 5& 4& 3& 5&4 &3 &4 &4 &5 & 4& 5& 3& 4\\ \hline
viii&3 & 4& 4& 5& 3& 4&4 &5 &4 &3 & 5& 4& 4& 5& 3& 4\\ \hline
ix & 3& 3& 5& 5& 3&4 & 3&4 &5 & 4&5 &4 & 4& 4& 4& 4\\ \hline
x & 3& 3& 5& 5& 4& 3& 4& 3& 4& 5& 4&5 & 4& 4& 4& 4\\ \hline
xi& 3& 4& 5& 6& 3& 4& 4& 5& 5& 4& 6&5 & 4& 5& 4& 5\\ \hline
xii & 3& 4& 5& 6& 4& 3& 5&4 & 4& 5&5 &6 & 4& 5& 4& 5\\ \hline
xiii& 3& 3& 5& 5&4 &4 &4 &4 &4 &4 &4 &4 &5 &5 &3 &3 \\ \hline
xiv& 3& 4& 5& 6&4 &4 &5 &5 &4 &4 &5 &5 & 5& 6& 3&4 \\ \hline
xv& 3& 3& 4& 4& 3& 3& 3& 3& 4&4 &4 &4 & 3& 3&4 &4 \\ \hline
xvi& 3& 4& 4& 5& 3&3 &4 &4 &4 &4 &5 &5 & 3& 4& 4& 5\\ \hline
\end{tabular}
\end{center}
\caption{The neutrino generation numbers for  combinations of zero-modes $\beta_i$ and $\gamma_j$.}
\label{tab:n-generation}
\end{table}

\clearpage

Table \ref{tab:possible-n} shows  the combination numbers of zero-modes $\beta_i$ and $\gamma_j$ 
for the neutrino generation numbers.
That implies that the three and four generations are statistically favored.
Their probabilities are 32\% and 42\%.

Similarly, Table \ref{tab:n-number-SS} shows the combination numbers of zero-modes $\beta_i$ and $\gamma_j$ 
for the neutrino generation numbers such that the neutrino sector has vanishing SS phases.
It means the combinations of $\beta_i$ and $\gamma_j$ corresponding to 
the SS phases, $(0,0)$ and $(0,0)$, $(1/2,1/2)$ and $(1/2,1/2)$,  $(1/2,0)$ and $(1/2,0)$,  $(0,1/2)$ and $(0,1/2)$.
Again, the three and four generations are statistically favored.
Their probabilities are about 42\% and 28\%.

\begin{table}[h]
\begin{center}
\begin{tabular}{|c|c|c|c|c|c|} \hline
neutrino generation number & 3 & 4 & 5 & 6 & 7 \\ \hline
combination number of $\beta$ and $\gamma$ & 81 & 108 & 54 & 12 &1 \\ \hline
\end{tabular}
\end{center}
\caption{The  combination numbers of zero-modes $\beta_i$ and $\gamma_j$  for the neutrino generation numbers.
The number of all the possible combinations  $\beta_i$ and $\gamma_j$  is equal to 256.}
\label{tab:possible-n}
\end{table}

\begin{table}[h]
\begin{center}
\begin{tabular}{|c|c|c|c|c|c|} \hline
neutrino generation number & 3 & 4 & 5 & 6 & 7 \\ \hline
combination number of $\beta$ and $\gamma$ & 27 & 18 & 12 & 6 &1 \\ \hline
\end{tabular}
\end{center}
\caption{The  combination numbers of zero-modes $\beta_i$ and $\gamma_j$  for the neutrino generation numbers 
 such that the neutrino sector has vanishing SS phases.
The number of all the possible combinations $\beta_i$ and $\gamma_j$   is equal to 64.}
\label{tab:n-number-SS}
\end{table}

\section{Majorana neutrino mass matrices}
\label{sec:Neutrino-mass}

Here we show explicitly Majorana neutrino mass matrices induced by 
D-brane instanton effects.
We restrict ourselves the three-generation models, where neutrinos correspond to zero-modes 
with vanishing SS phases.

\subsection{Neutrino mass terms}
\label{subsec: neutrino mass}
In the previous sections, we have shown possible D-brane configurations, where 
Majorana neutrino mass terms can be induced by 
D-brane instanton effects.
All of the possible effects can contribute on neutrino mass terms for a fixed configuration of 
D-branes D$_{N1}$ and D$_{N2}$.

When the neutrinos appear as three $\mathbb{Z}_2$ even zero-modes corresponding to 
the magnetic flux $M_N=4$ and 5, D-brane instanton effects to generate Majorana neutrino masses 
are unique as shown in subsections \ref{subsec:2-2-4} and \ref{subsec:2-3-5}.
On the other hand, when neutrinos appear as three $\mathbb{Z}_2$ odd zero-modes corresponding to 
the magnetic flux $M_N=7$ and 8, 
the total mass terms are written by linear combinations with the factor $e^{-S_{cl}(D_{inst},M_{inst})}$.
When neutrinos correspond to 
the magnetic flux $M_N=8$, 
their mass matrix is written by 
\begin{eqnarray}
\label{eq: full mass}
M_{ab}&=&e^{-S_{cl}(D_{inst},M_{inst}^{(2-6-8)})}m_{ab}^{(2-6-8)} 
+\sum_{(\alpha_1,\alpha_\tau)}e^{-S_{cl}(D_{inst},M_{inst}^{(3-5-8)})}m_{ab}^{(3-5-8)(\alpha_1,\alpha_\tau)}\nonumber \\
&&+
\sum_{(\alpha_1,\alpha_\tau)}e^{-S_{cl}(D_{inst},M_{inst}^{(4-4-8)})}m_{ab}^{(4-4-8)(\alpha_1,\alpha_\tau)}.
\end{eqnarray}
Here, $M_{inst}^{(2-6-8)}$ denotes the magnetic flux on the D-brane instanton to realize 
the case with $(M_\beta,M_\gamma,M_N)=(2,6,8)$, which is denoted  in short by 
the 2-6-8 case.
Other notations such as $M_{inst}^{(3-5-8)}$ and $M_{inst}^{(4-4-8)}$ 
 have similar meaning. Note that we can shift the magnetic fluxes on D$_{N1}$, D$_{N2}$ and D$_{inst}$-branes by constant 
to make their differences invariant and realize the 2-6-8 case.
For example, when we chose magnetic fluxes on D$_{N1}$ and D$_{N2}$-branes such that 
$M_{inst}^{(2-6-8)}=0$ and $M_{inst}^{(3-5-8)}, M_{inst}^{(4-4-8)} \neq 0$, 
the first term is dominant, and other terms are exponentially suppressed, 
\begin{eqnarray}
\label{eq: dominant mass}
M_{ab} \approx e^{-S_{cl}(D_{inst},0)}m_{ab}^{(2-6-8)} .
\end{eqnarray}

Similarly, when $M_N=7$, the mass matrix is written by
\begin{eqnarray}
M_{ab}=e^{-S_{cl}(D_{inst},M_{inst}^{(2-5-7)})}m_{ab}^{(2-5-7)} 
+\sum_{(\alpha_1,\alpha_\tau)}e^{-S_{cl}(D_{inst},M_{inst}^{(3-4-7)})}.
\end{eqnarray}
Under the choice $M_{inst}^{(2-5-7)}=0$ and $M_{inst}^{(3-4-7)} \neq 0$, we can approximate 
\begin{equation}
M_{ab} \approx e^{-S_{cl}(D_{inst},0)}m_{ab}^{(2-5-7)} . \label{evM7}
\end{equation}

\subsection{Neutrino sector with $M_N=4$}
\label{subsec:2-2-4}
Here, we study the mass matrix of the neutrino sector with $M_N=4$.
The only possibility is the 2-2-4 case which is constructed as follows.
Two independent zero-modes for both $\beta_i$ and $\gamma_j$ are obtained by taking even wavefunctions under the magnetic flux, $M_{\beta}=M_{\gamma}=2$. For the neutrino sector, three independent zero-modes are obtained by taking even wavefunctions under the magnetic flux, $M_N=4$. 
Wavefunctions for $\beta_i$ and $\gamma_j$ are given by
\begin{align}
\begin{aligned}
&\psi_{T^2/\mathbb{Z}_2^+}^{(0,0),2}(z) = \psi^{(0,0), 2}_{T^2}(z), \\
&\psi_{{T^2/\mathbb{Z}_2^+}}^{(1,0),2}(z) = \psi^{(1,0), 2}_{T^2}(z).
\end{aligned}
\end{align}
Wavefunctions for the neutrino sector are given by
\begin{align}
\begin{aligned}
&\psi_{{T^2/\mathbb{Z}_2^+}}^{(0,0),4}(z) = \psi^{(0,0),4}_{T^2}(z), \\
&\psi_{{T^2/\mathbb{Z}_2^+}}^{(1,0),4}(z) = \frac{1}{\sqrt{2}}(\psi^{(1,0),4}_{T^2}(z) + \psi^{(3,0),4}_{T^2}(z)), \\
&\psi_{{T^2/\mathbb{Z}_2^+}}^{(2,0),4}(z) = \psi^{(2,0),4}_{T^2}(z).
\end{aligned}
\end{align}
From the above wavefunctions, the $d$ matrices can be computed to give
\begin{align}
\begin{aligned}
d_1 &= c_{(2-2-4)}
\begin{pmatrix}
\eta^{(16)}_0 + \eta^{(16)}_8 & 0 \\
0 & \eta^{(16)}_4 + \eta^{(16)}_{12} \\
\end{pmatrix}, \\ 
d_2 &=  c_{(2-2-4)}\sqrt{2} 
\begin{pmatrix}
0 & \eta^{(16)}_2 + \eta^{(16)}_{10} \\
\eta^{(16)}_2 + \eta^{(16)}_{10} & 0
\end{pmatrix}, \\  
d_3 &= c_{(2-2-4)}
\begin{pmatrix}
\eta^{(16)}_4 + \eta^{(16)}_{12} & 0 \\
0 & \eta^{(16)}_0 + \eta^{(16)}_8 
\end{pmatrix},
\end{aligned}
\end{align}
where we have defined
\begin{equation}
\eta^{(n)}_N = \vartheta \biggl[\begin{array}{c} \frac{N}{n}\\      0 \end{array} \biggr](0, n\tau).
\end{equation}
The explicit form of the overall constant factor $c_{(2-2-4)}$ is shown by Eq. ($\ref{eq: overall constant}$).
The mass matrix is then given by
\begin{eqnarray}
\bm{m}^{(2-2-4)}= c^2_{(2-2-4)} 
\begin{pmatrix}
X_3 & 0 & X_1 \\
0 & -\sqrt{2} X_2 & 0 \\
X_1 & 0 & X_3 
\end{pmatrix},
\end{eqnarray} 
where $X_i \ (i=1,2,3)$ are defined as
\begin{align}
\begin{aligned}
X_1 &= (\eta^{(16)}_0 + \eta^{(16)}_8)^2 + (\eta^{(16)}_4 + \eta^{(16)}_{12})^2,\\
X_2 &= \frac{1}{\sqrt{2}} \left( (\eta^{(16)}_2 + \eta^{(16)}_{10} ) + (\eta^{(16)}_6 + \eta^{(16)}_{14} ) \right)^2 , \\ 
X_3 &= 2(\eta^{(16)}_0 + \eta^{(16)}_8)(\eta^{(16)}_4 + \eta^{(16)}_{12}).
\end{aligned}
\end{align}

Next, we investigate the modular transformation behavior of $\bm{m}^{(2-2-4)}$.
Under the $S$-transformation, $\tau \rightarrow -\frac{1}{\tau}$, $X_i \  ( i = 1,2,3)$ are transformed as,
\begin{equation}
\begin{pmatrix}
X_1 \\
X_2 \\
X_3 
\end{pmatrix}
\xrightarrow{S} (-\tau) \frac{i}{2}
\begin{pmatrix}
1 & \sqrt2 & 1 \\
\sqrt2 & 0 & -\sqrt2 \\
1 & -\sqrt2 & 1
\end{pmatrix}
\begin{pmatrix}
X_1 \\
X_2 \\
X_3 
\end{pmatrix}.
\label{224XS}
\end{equation}
Under the $T$-transformation, $\tau \rightarrow \tau + 1$,  \begin{equation}
\begin{pmatrix}
X_1 \\
X_2 \\
X_3 
\end{pmatrix}
\xrightarrow{T}  
\begin{pmatrix}
1 & 0 & 0 \\
0 & i & 0 \\
0 & 0 & -1
\end{pmatrix}
\begin{pmatrix}
X_1 \\
X_2 \\
X_3 
\end{pmatrix}.
\label{224XT}
\end{equation}
From the above results, it can be seen that $X_1, X_2, {\rm and}\  X_3$ are modular forms of the weight 1 and behave as a triplet of the group $S'_4$~\footnote{$\Gamma'_4 \simeq S'_4 \simeq \Delta'(24)$ is the double covering group of $\Gamma_4 \simeq S_4 \simeq \Delta(24)$.}.

\begin{center}
\begin{figure}[H]
\begin{minipage}{8cm}
\includegraphics[bb=0 0 700 625,width=5cm]{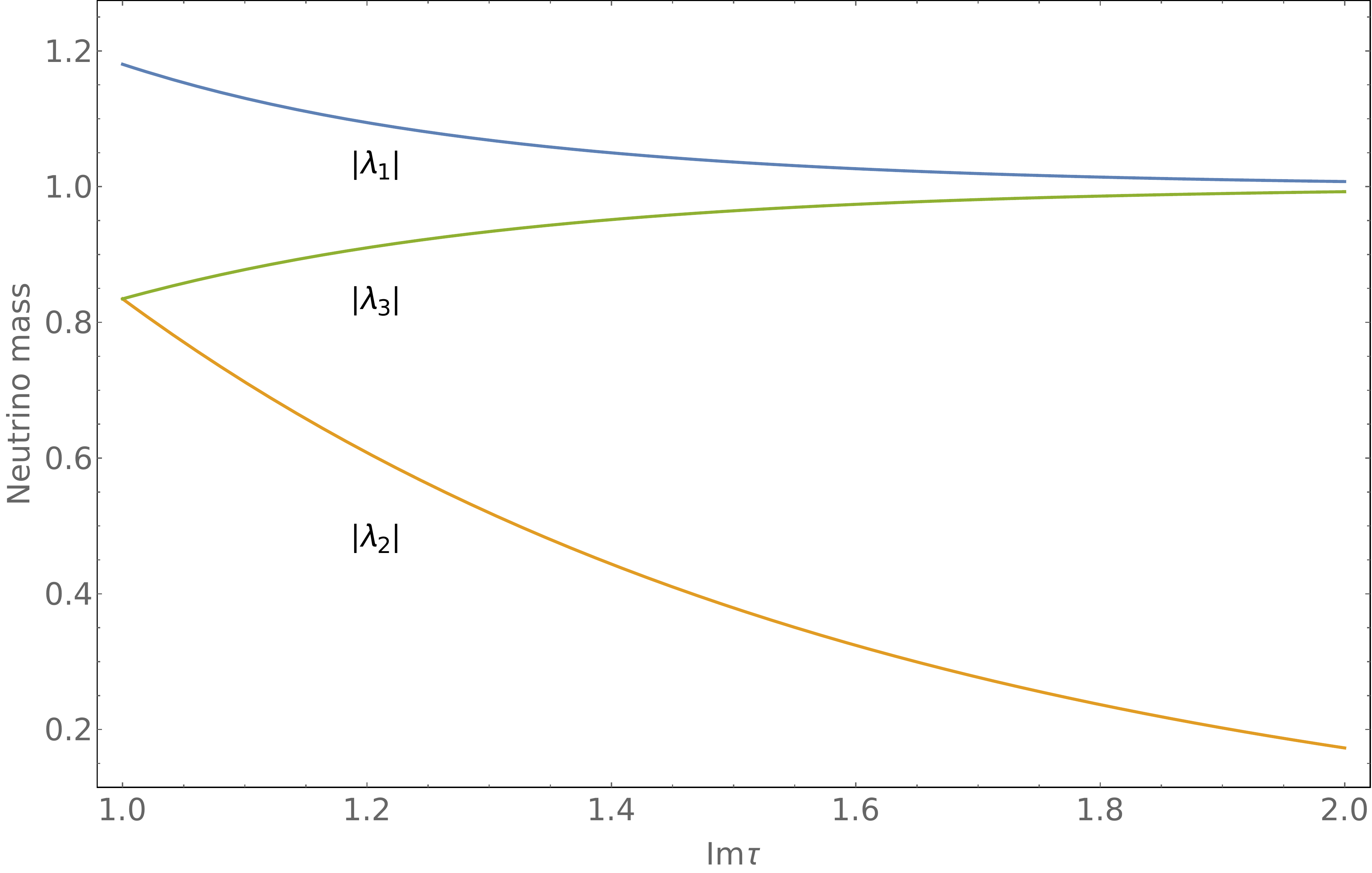}
\caption{${\rm Im}\tau$ dependence $(1 \leq {\rm Im}\tau \leq 2)$ of the absolute values of the mass eigenvalues in the 2-2-4 case at ${\rm Re}\tau=0$.}
\label{224Retau0}
\end{minipage}
\hfill
\begin{minipage}{8cm}
\includegraphics[bb=0 0 750 675,width=5cm]{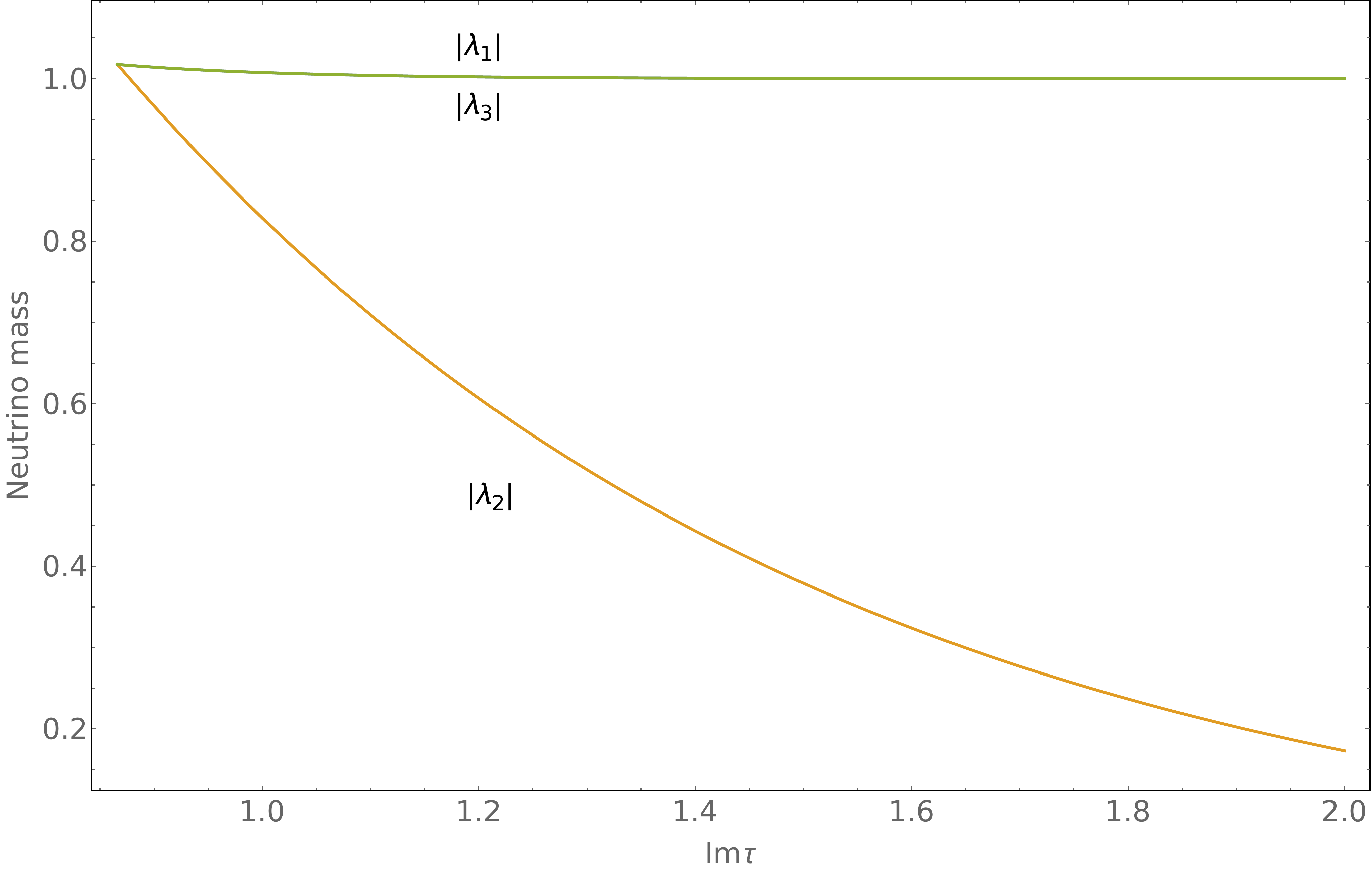}
\caption{${\rm Im}\tau$ dependence $(\sqrt{3}/2 \leq {\rm Im}\tau \leq 2)$ of the absolute values of the mass eigenvalues in the 2-2-4 case at ${\rm Re}\tau=1/2$.}
\label{224Retau1/2}
\end{minipage}
\end{figure}
\end{center}

Eigenvalues of $\bm{m}^{(2-2-4)}$ are given by
\begin{align}
\begin{aligned}
\lambda_1 &= c^2_{(2-2-4)}(X_1 + X_3) = c^2_{(2-2-4)} \left( (\eta^{(16)}_0 + \eta^{(16)}_8) + (\eta^{(16)}_4 + \eta^{(16)}_{12}) \right)^2, \\
\lambda_2 &= -c^2_{(2-2-4)}\sqrt{2} X_2 = - c^2_{(2-2-4)} \left( (\eta^{(16)}_2 + \eta^{(16)}_{10} ) + (\eta^{(16)}_6 + \eta^{(16)}_{14} ) \right)^2, \\
\lambda_3 &= c^2_{(2-2-4)}(X_3 - X_1) = - c^2_{(2-2-4)} \left( (\eta^{(16)}_0 + \eta^{(16)}_8) - (\eta^{(16)}_4 + \eta^{(16)}_{12}) \right)^2 .
\end{aligned}
\label{ME224}
\end{align}
They can be approximated at large ${\rm Im}\tau$ as
\begin{align}
\begin{aligned}
\lambda_1 
&\approx c^2_{(2-2-4)} (\eta^{(16)}_0)^2 \approx c^2_{(2-2-4)} ( 1+ \cdots), \\
\lambda_2 
&\approx - c^2_{(2-2-4)} 4 (\eta^{(16)}_2)^2 \approx - 4 c^2_{(2-2-4)} ( e^{\frac{\pi i \tau}{2}} + \cdots), \\
\lambda_3 
&\approx - c^2_{(2-2-4)} (\eta^{(16)}_0)^2 \approx - c^2_{(2-2-4)} ( 1+ \cdots) .
\end{aligned}
\label{ME224'}
\end{align}
The diagonalizing matrix $P$ for $\forall \tau$ satisfying $P^{\mathrm T} \bm{m}^{(2-2-4)} P = {\rm diag}(\lambda_1, \lambda_2, \lambda_3)$ is 
\begin{equation}
P = 
\begin{pmatrix}
\frac{1}{\sqrt2} & 0 & \frac{-1}{\sqrt2} \\
0 & 1 & 0 \\
\frac{1}{\sqrt2} & 0 & \frac{1}{\sqrt2} 
\end{pmatrix},
\label{224P}
\end{equation}
showing that the mixing angle is $45^{\circ}$. Note that the mass eigenstates become $\mathbb{Z}_2$-twisted and $\mathbb{Z}_2$-shifted eigenstates\footnote{They can be defined in $M \in 4\mathbb{Z}$. See in detail Refs.~\cite{Kikuchi:2020frp,Kikuchi:2020nxn}.}.
Figures \ref{224Retau0} and \ref{224Retau1/2} show the ${\rm Im}\tau$ dependence $(\sqrt{1-({\rm Re}\tau)^2} \leq {\rm Im}\tau \leq 2)$ of the absolute values of the mass eigenvalues $\lambda_i$ ($i=1,2,3$) in Eq.~(\ref{ME224}) at ${\rm Re}\tau=0$ and ${\rm Re}\tau=1/2$, respectively.
Here, we set $c_{(2-2-4)}=1$ for simplicity.

There are four interesting features in Figures \ref{224Retau0} and \ref{224Retau1/2}.
First, we can find $|\lambda_2|=|\lambda_3|$ at $\tau=i$. 
It can be explained by considering that the point $\tau=i$ is invariant under $S$-transformation.
In other words, each mass eigenvalue $\lambda_k$ ($k=1,2,3$) at $\tau=i$ must be invariant under $S$-transformation.
However, from Eqs~(\ref{224XS}) and (\ref{224P}), $\lambda_k$ ($k=1,2,3$) at $\tau=i$ are transformed under $S$-transformation as,
\begin{equation}
\begin{pmatrix}
\lambda_1 \\
\lambda_2 \\
\lambda_3 
\end{pmatrix}
\xrightarrow{S}
\begin{pmatrix}
1 & 0 & 0 \\
0 & 0 & 1 \\
0 & 1 & 0
\end{pmatrix}
\begin{pmatrix}
\lambda_1 \\
\lambda_2 \\
\lambda_3 
\end{pmatrix}.
\label{224MES}
\end{equation}
Thus, in order for $\lambda_k$ ($k=1,2,3$) at $\tau=i$ to be invariant under $S$-transformation, it should be required that $\lambda_2=\lambda_3$ at $\tau=i$.
Actually, by using the following relation,
\begin{align}
\eta_{N}^{(n)}(-1/\tau) = \sqrt{\frac{-i\tau}{n}} \sum_{N'=0}^{n-1} e^{2\pi i\frac{NN'}{n}} \eta_{N'}^{(n)}(\tau),
\end{align}
we can check
\begin{align}
(\eta^{(16)}_0 + \eta^{(16)}_8) - (\eta^{(16)}_4 + \eta^{(16)}_{12}) = (\eta^{(16)}_2 + \eta^{(16)}_{10}) + (\eta^{(16)}_6 + \eta^{(16)}_{14}), \quad (\tau=i)
\end{align}
and then $\lambda_2=\lambda_3$ is satisfied at $\tau=i$.

Second, we can find $|\lambda_1| \simeq |\lambda_3|$ and $|\lambda_2| \rightarrow 0$ at $\tau \rightarrow i \infty$. 
It can be explained by considering that the limit $\tau=i \infty$ is invariant under $T$-transformation.
In other words, each mass eigenvalue $\lambda_k$ ($k=1,2,3$) at $\tau=i \infty$ must be invariant under $T$-transformation.
However, from Eqs~(\ref{224XT}) and (\ref{224P}), $\lambda_k$ ($k=1,2,3$) are transformed under $T$-transformation as,
\begin{equation}
\begin{pmatrix}
\lambda_1 \\
\lambda_2 \\
\lambda_3 
\end{pmatrix}
\xrightarrow{T}
\begin{pmatrix}
0 & 0 & -1 \\
0 & i & 0 \\
-1 & 0 & 0
\end{pmatrix}
\begin{pmatrix}
\lambda_1 \\
\lambda_2 \\
\lambda_3 
\end{pmatrix}.
\label{224MET}
\end{equation}
Thus, in order for $\lambda_k$ ($k=1,2,3$) at $\tau=i \infty$ to be invariant under $T$-transformation, it should be required that $\lambda_1=-\lambda_3$ and $\lambda_2=i\lambda_2 \Leftrightarrow \lambda_2=0$ at $\tau=i \infty$.
Actually, $\lambda_k$ ($k=1,2,3$) in Eq.~(\ref{ME224'}) at $\tau \rightarrow i \infty$ are estimated as $\lambda_1 \simeq -\lambda_3 \rightarrow 1$ and $\lambda_2 \rightarrow 0$.

Third, we can find $|\lambda_1|=|\lambda_3|$ at $\tau=\frac{1}{2}+i{\rm Im}\tau$.
It can be explained by considering that $\tau=\frac{1}{2}+i{\rm Im}\tau$ is invariant under $T$-transformation and  $CP$-transformation, where $CP: \tau \rightarrow -\bar{\tau}$.
In other words, each mass eigenvalue $\lambda_k$ ($k=1,2,3$) at $\tau=\frac{1}{2}+i{\rm Im}\tau$ must be invariant under $T \cdot CP$-transformation.
Here, we can easily check that
\begin{equation}
CP: \lambda_k(\tau) \rightarrow \lambda_k(-\bar{\tau}) = \lambda_k^{\ast}(\tau), \ (k=1,2,3). \label{224MECP}
\end{equation}
Then, $\lambda_k$ ($k=1,2,3$) are transformed under $T \cdot CP$-transformation as,
\begin{equation}
\begin{pmatrix}
\lambda_1 \\
\lambda_2 \\
\lambda_3 
\end{pmatrix}
\xrightarrow{T \cdot CP}
\begin{pmatrix}
0 & 0 & -1 \\
0 & i & 0 \\
-1 & 0 & 0
\end{pmatrix}
\begin{pmatrix}
\lambda_1^{\ast} \\
\lambda_2^{\ast} \\
\lambda_3^{\ast} 
\end{pmatrix}.
\label{224MET}
\end{equation}
Thus, in order for $\lambda_k$ ($k=1,2,3$) at $\tau=\frac{1}{2}+i{\rm Im}\tau$ to be invariant under $T \cdot CP$-transformation, it should be required that $\lambda_1=-\lambda_3^{\ast}$ and $\lambda_2=i\lambda_2^{\ast}$ at $\tau=\frac{1}{2}+i{\rm Im}\tau$.
Actually, $\lambda_k$ ($k=1,2,3$) in Eq.~(\ref{ME224}) at $\tau=\frac{1}{2}+i{\rm Im}\tau$ are expressed as,
\begin{align}
\begin{aligned}
\lambda_1 &= \left( (|\eta^{(16)}_0| + |\eta^{(16)}_8|) + i (|\eta^{(16)}_4| + |\eta^{(16)}_{12}|) \right)^2, \\
\lambda_2 &= - 4 e^{\frac{\pi i}{4}} \left(|\eta^{(16)}_2| - |\eta^{(16)}_{10}| \right)^2, \\
\lambda_3 &= - \left( (|\eta^{(16)}_0| + |\eta^{(16)}_8|) - i (|\eta^{(16)}_4| + |\eta^{(16)}_{12}|) \right)^2,
\end{aligned}
\label{ME224third}
\end{align}
and then they satisfy $\lambda_1=-\lambda_3^{\ast}$ and $\lambda_2=i\lambda_2^{\ast}$.

Fourth, we can find $|\lambda_1|=|\lambda_2|=|\lambda_3|$ at $\tau=\frac{1}{2}+\frac{\sqrt{3}}{2}i$. 
It can be explained by considering that the point $\tau=\frac{1}{2}+\frac{\sqrt{3}}{2}i$ is invariant under $ST^{-1}$-transformation.
In other words, each mass eigenvalue $\lambda_k$ ($k=1,2,3$) at $\tau=\frac{1}{2}+\frac{\sqrt{3}}{2}i$ must be invariant under $ST^{-1}$-transformation.
However, $\lambda_k$ ($k=1,2,3$) at $\tau=\frac{1}{2}+\frac{\sqrt{3}}{2}i$ are transformed under $ST^{-1}$-transformation as,
\begin{equation}
\begin{pmatrix}
\lambda_1 \\
\lambda_2 \\
\lambda_3 
\end{pmatrix}
\xrightarrow{ST^{-1}}
\begin{pmatrix}
0 & 0 & e^{-\frac{5\pi i}{6}} \\
e^{-\frac{5\pi i}{6}} & 0 & 0 \\
0 & e^{-\frac{\pi i}{3}} & 0
\end{pmatrix}
\begin{pmatrix}
\lambda_1 \\
\lambda_2 \\
\lambda_3 
\end{pmatrix}.
\label{224MES}
\end{equation}
Thus, in order for $\lambda_k$ ($k=1,2,3$) at $\tau=\frac{1}{2}+\frac{\sqrt{3}}{2}i$ to be invariant under $ST^{-1}$-transformation, it should be required that $\lambda_1=e^{\frac{5\pi i}{6}}\lambda_2=e^{-\frac{5\pi i}{6}}\lambda_3$ at $\tau=\frac{1}{2}+\frac{\sqrt{3}}{2}i$.
Actually, this can be also checked explicitly by considering $\tau=\frac{1}{2}+\frac{\sqrt{3}}{2}i$ is invariant under $ST^{-1}$-transformation.
Note that $\lambda_1=-\lambda_3^{\ast}$ and $\lambda_2=i\lambda_2^{\ast}$ should be also satisfied and actually they are consistent with $\lambda_1=e^{\frac{5\pi i}{6}}\lambda_2=e^{-\frac{5\pi i}{6}}\lambda_3$.

\subsection{Neutrino sector with $M_N=8$}

Here, we study mass matrix of the neutrino sector with $M_N=8$.
For this neutrino sector, the mass terms can be induced by several cases, 
i.e. the 2-6-8 case,  the 3-5-8 case, and the 4-4-8 case.

\subsubsection{2-6-8 Case}
In the 2-6-8 case, we take even wavefunctions under $M_{\beta}=2$ and odd wavefunctions under $M_{\gamma}=6$, for $\beta_i$ and $\gamma_j$ respectively. For the neutrino sector, we take odd wavefunctions under $M_N=8$. 
The $d$ matrices are given by
\begin{eqnarray}
d_1 = c_{(2-6-8)}
\begin{pmatrix}
A_1& 0 \\
0 & A_3 \\
\end{pmatrix}
, \ 
d_2 =  c_{(2-6-8)}
\begin{pmatrix}
0 & A_2 \\
A_2 & 0
\end{pmatrix}
, \ 
d_3 = c_{(2-6-8)}
\begin{pmatrix}
A_3 & 0 \\
0 & A_1 
\end{pmatrix}.
\end{eqnarray}
Here, $A_i \ (i=1,2,3)$ are defined as
\begin{align}
\begin{aligned}
A_1 &= (\eta^{(96)}_2 - \eta^{(96)}_{14}) - (\eta^{(96)}_{34} - \eta^{(96)}_{46}), \\
A_2 &= (\eta^{(96)}_4 - \eta^{(96)}_{28}) - (\eta^{(96)}_{20} - \eta^{(96)}_{44}), \\
A_3 &= (\eta^{(96)}_{10} - \eta^{(96)}_{22}) - (\eta^{(96)}_{26} - \eta^{(96)}_{38}). 
\end{aligned}
\end{align}
The mass matrix is then given by
\begin{eqnarray}
\label{eq: 2-6-8}
\bm{m}^{(2-6-8)}=c^2_{(2-6-8)}
\begin{pmatrix}
X_3 & 0 & X_1 \\
0 & -\sqrt2 X_2 & 0 \\
X_1 & 0 & X_3      
\end{pmatrix},
\end{eqnarray} 
where $X_i \ (i=1,2,3)$ are defined as
\begin{align}
\begin{aligned}
X_1 &= (A_1)^2 + (A_3)^2, \\
X_2 &= \sqrt2(A_2)^2, \\
X_3 &= 2A_1 A_3.
\end{aligned}
\end{align}

Next, let us investigate the modular transformation behavior of the mass matrix, $\bm{m}^{(2-6-8)}$.
Under the $S$-transformation, $\tau \rightarrow -\frac{1}{\tau}$, $X_i \  ( i = 1,2,3)$ are transformed as,
\begin{equation}
\begin{pmatrix}
X_1 \\
X_2 \\
X_3 
\end{pmatrix}
\xrightarrow{S} (-\tau) \frac{i}{2}
\begin{pmatrix}
1 & \sqrt2 & 1 \\
\sqrt2 & 0 & -\sqrt2 \\
1 & -\sqrt2 & 1
\end{pmatrix}
\begin{pmatrix}
X_1 \\
X_2 \\
X_3 
\end{pmatrix}.
\end{equation}
Under the $T$-transformation, $\tau \rightarrow \tau + 1$,  \begin{equation}
\begin{pmatrix}
X_1 \\
X_2 \\
X_3 
\end{pmatrix}
\xrightarrow{T}  
e^{\frac{i \pi}{12}} 
\begin{pmatrix}
1 & 0 & 0 \\
0 & e^{\frac{i \pi}{4}} & 0 \\
0 & 0 & -1 
\end{pmatrix}
\begin{pmatrix}
X_1 \\
X_2 \\
X_3 
\end{pmatrix}.
\end{equation}
From the above results, it can be seen that $X_1, X_2, {\rm and}\  X_3$ are modular forms of the weight 1 and behave as a triplet of the group $\Delta'(96) \times \mathbb{Z}_3$~\footnote{$\Delta'(96) \simeq (\mathbb{Z}_4 \times \mathbb{Z}'_4) \rtimes \mathbb{Z}_3 \rtimes \mathbb{Z}_4$, which is the double covering group of $\Delta(96) \subset \Gamma_8$, is a subgroup of $\Gamma'_8$~\cite{Kikuchi:2021ogn}.}.

\subsubsection{3-5-8 Case}
In the 3-5-8 case, there are four possible variations with different SS phases, (0,0), (1/2,0), (0,1/2), and (1/2,1/2).
The sum of four mass matrices\footnote{In particular, $m^{(3-5-8)}_{12}$ and $m^{(3-5-8)}_{23}$ (as well as $m^{(3-5-8)}_{21}$ and $m^{(3-5-8)}_{32}$) are canceled each other by those components of $\bm{m}^{(3-5-8)(\alpha_1,0)}$ and $\bm{m}^{(3-5-8)(\alpha_1,1/2)}$ with $\alpha_1=0,1/2$.} in equal ratio is given by
\begin{equation}
\label{eq: 3-5-8}
\begin{split}
\bm{m}^{(3-5-8)}=\bm{m}^{(3-5-8)(0,0)} +& \bm{m}^{(3-5-8)(1/2,0)} + \bm{m}^{(3-5-8)(0,1/2)} + \bm{m}^{(3-5-8)(1/2,1/2)} 
\\ &= c^2_{(3-5-8)}
\begin{pmatrix}
Y_3 & 0 & Y_1 \\
0 & -\sqrt2 Y_2 & 0 \\
Y_1 & 0 & Y_3
\end{pmatrix}.
\end{split}
\end{equation}
Here, $Y_1, Y_2$ and $Y_3$ are defined as
\begin{align}
\begin{aligned}
Y_1 &= 2\sqrt2 (D_{9} \tilde{D}_{37} + D_{33} \tilde{D}_{-11} + D_{-3} \tilde{D}_{1} + D_{21} \tilde{D}_{-47}),\\
Y_2 &= 4 (E_{-6} \tilde{E}_{-2} - E_{42}\tilde{E}_{34}), \\
Y_3 &= 2\sqrt2 (D_{9} \tilde{D}_{-47} + D_{33} \tilde{D}_{1} + D_{21} \tilde{D}_{37} + D_{-3} \tilde{D}_{-11}),
\end{aligned}
\end{align}
where $D_N, E_N, \tilde{D}_N, {\rm and}\  \tilde{E}_N$ are given by
\begin{align}
\begin{aligned}
D_N &= \eta^{(120)}_N - \eta^{(120)}_{N+30}, \\
E_N &= \eta^{(120)}_N - \eta^{(120)}_{N+60}, \\
\tilde{D}_N &= D_N + D_{N+40}, \\
\tilde{E}_N &= E_N + E_{N+40}. 
\end{aligned}
\end{align}
Under the modular transformation, $Y_1, Y_2$, and $Y_3$ are modular forms of the weight 1 and behave as a triplet of the group $\Delta'(96) \times \mathbb{Z}_3$.

\subsubsection{4-4-8 Case}
In the 4-4-8 case, there are three possible variations with different SS phases, (1/2,0), (0,1/2), and (1/2,1/2).
The sum of three mass matrices in equal ratio is given by
\begin{equation}
\label{eq: 4-4-8}
\begin{split}
\bm{m}^{(4-4-8)}=&\bm{m}^{(4-4-8)(1/2,0)} + \bm{m}^{(4-4-8)(0,1/2)} + \bm{m}^{(4-4-8)(1/2,1/2)} 
\\=&  c_{(4-4-8)}^2
\begin{pmatrix}
\sqrt2 Z_6 - Z_3 & 0 & Z_1 + \sqrt2 (Z_2 + Z_4) \\
0 & -2 (Z_1 + Z_5) & 0 \\
Z_1 + \sqrt2 (Z_2 + Z_4) & 0 & \sqrt2 Z_6 - Z_3
\end{pmatrix}.
\end{split}
\end{equation}
Here, $Z_i \ (i=1,2,3,4,5,6)$ are defined as
\begin{align}
\begin{aligned}
Z_1 &= (B_1)^2 + (B_3)^2, \qquad
Z_2 = \frac{1}{2\sqrt2} (B_0 - B_4)^2, \qquad
Z_3 = -2B_1B_3,  \\
Z_4 &= \frac{1}{2\sqrt2} (B_0 + B_4)^2 , \qquad
Z_5 = B_2 (B_0 + B_4) , \qquad
Z_6 = \sqrt2 (B_2)^2,
\end{aligned}
\end{align}
where $B_N$ is given by
\begin{equation}
B_N = \eta^{(128)}_{4N} + \eta^{(128)}_{4N+32} + \eta^{(128)}_{4N+64} + \eta^{(128)}_{4N+96}.
\end{equation} 
Under the modular transformation, $Z_i (i=1,2,3,4,5,6)$ are modular forms of the weight 1. It can be verified that $Z_i (i=1,2,3)$ behave as a triplet of the group $\Delta'(96)$. On the other hand, $Z_i (i=4,5,6)$ behave as a triplet of the group $S_4' \subset \Delta'(96)$.

\subsubsection{Total mass matrix and its eigenvalues}
Recall that the total mass matrix $\bm{M}^{(M_N=8)}$ is given by the  sum of $\bm{m}^{(2-6-8)},\ \bm{m}^{(4-4-8)}$, and $\bm{m}^{(3-5-8)}$ which are weighted by the factor $e^{-S_{cl}(D_{inst},M_{inst})}$ as shown in Eq. ($\ref{eq: full mass}$). In fact, $\bm{M}^{(M_N=8)}$ is diagonalized by the bimaximal mixing matrix,
\begin{equation}
\label{eq: bimaximal}
P = 
\begin{pmatrix}
\frac{1}{\sqrt2} & 0 & \frac{-1}{\sqrt2} \\
0 & 1 & 0 \\
\frac{1}{\sqrt2} & 0 & \frac{1}{\sqrt2} 
\end{pmatrix},
\end{equation}
meaning that we have $P^{\rm T}\bm{M}^{(M_N=8)} P  = {\rm diag}(\lambda_{1}^{(M_N = 8)} , \lambda_{2}^{(M_N = 8)} , \lambda_{3}^{(M_N = 8)} )$ independent of the factor $e^{-S_{cl}(D_{inst},M_{inst})}$ as well as the modulus $\tau$. This follows from the fact that Eq.~$(\ref{eq: 2-6-8})$, Eq.~$(\ref{eq: 3-5-8})$, and Eq.~$(\ref{eq: 4-4-8})$ are all separately diagonalized by Eq.~$ (\ref{eq: bimaximal})$ for $\forall \tau$. Note that the mass eigenstates become $\mathbb{Z}_2$-twisted and $\mathbb{Z}_2$-shifted eigenstates.
Let us denote the eigenvalues of $\bm{m}^{(2-6-8)}$ by $\lambda_{i}^{(2-6-8)} \ (i = 1,2,3)$. We understand $\lambda_{i}^{(3-5-8)}$ and $\lambda_{i}^{(4-4-8)}$ in the same way. Then, eigenvalues  $\lambda_{i}^{(M_N = 8)}$ are given by
\begin{eqnarray}
\lambda_{i}^{(M_N = 8)} &= &e^{-S_{cl}(D_{inst},M_{inst}^{(2-6-8)})}\lambda_{i}^{(2-6-8)} 
+e^{-S_{cl}(D_{inst},M_{inst}^{(3-5-8)})}\lambda_{i}^{(3-5-8)}\nonumber \\
&&+e^{-S_{cl}(D_{inst},M_{inst}^{(4-4-8)})}\lambda_{i}^{(4-4-8)}, \qquad   
\end{eqnarray}
where
\begin{align}
\begin{aligned}
\lambda_{1}^{(2-6-8)} &= c^2_{(2-6-8)} (X_1 + X_3) = c^2_{(2-6-8)} (A_1+A_3)^2, \\
\lambda_{2}^{(2-6-8)} &= -c^2_{(2-6-8)} \sqrt{2} X_2 = -c^2_{(2-6-8)} 2(A_2)^2, \\
\lambda_{3}^{(2-6-8)} &= c^2_{(2-6-8)} (X_3 - X_1) = -c^2_{(2-6-8)} (A_1-A_3)^2, \\
\label{ME268}
\end{aligned}
\end{align}
\begin{align}
\begin{aligned}
\lambda_{1}^{(3-5-8)} &= c^2_{(3-5-8)}(Y_1 + Y_3) \\
&= c^2_{(3-5-8)} 2\sqrt{2}\left( (D_{9}+D_{21})(\tilde{D}_{37}+\tilde{D}_{-47})+(D_{33}+D_{-3})(\tilde{D}_{-11}+\tilde{D}_1) \right), \\
\lambda_{2}^{(3-5-8)} &= - c^2_{(3-5-8)} \sqrt2 Y_2 \\
&= - c^2_{(3-5-8)} 4\sqrt{2}(E_{-6} \tilde{E}_{-2} - E_{42}\tilde{E}_{34}), \\
\lambda_{3}^{(3-5-8)} &= c^2_{(3-5-8)}(Y_3 -Y_1) \\
&= -c^2_{(3-5-8)} 2\sqrt{2}\left( (D_{9}-D_{21})(\tilde{D}_{37}-\tilde{D}_{-47})+(D_{33}-D_{-3})(\tilde{D}_{-11}-\tilde{D}_1) \right), \\
\label{ME358}
\end{aligned}
\end{align}
\begin{align}
\begin{aligned}
\lambda_{1}^{(4-4-8)} &= c^2_{(4-4-8)} \left(\sqrt2 ( Z_2 + Z_4 + Z_6 ) + (Z_1 - Z_3)\right) \\
&= c^2_{(4-4-8)} \left((B_0^2+B_4^2+2B_2^2) + (B_1+B_3)^2\right), \\
\lambda_{2}^{(4-4-8)} &= -c^2_{(4-4-8)} 2(Z_1 + Z_5) \\
&= -c^2_{(4-4-8)} 2\left( (B_1)^2 + (B_3)^2 + B_2 (B_0 + B_4) \right), \\
\lambda_{3}^{(4-4-8)} &= c^2_{(4-4-8)} \left(\sqrt2(Z_6 - Z_2 -Z_4) - (Z_1 + Z_3)\right) \\
&= -c^2_{(4-4-8)} \left((B_0^2+B_4^2-2B_2^2) + (B_1-B_3)^2\right). \\
\label{ME448}
\end{aligned}
\end{align}
They can be approximated by 
\begin{align}
\begin{aligned}
\lambda_{1}^{(2-6-8)} 
&\approx c^2_{(2-6-8)} (\eta^{(96)}_2)^2 \approx  c^2_{(2-6-8)} (e^{\frac{\pi i \tau}{12}} + \cdots), \\
\lambda_{2}^{(2-6-8)} 
&\approx -c^2_{(2-6-8)} 2(\eta^{(96)}_4)^2 \approx -2c^2_{(2-6-8)} (e^{\frac{\pi i \tau}{3}} + \cdots), \\
\lambda_{3}^{(2-6-8)} 
&\approx - c^2_{(2-6-8)} (\eta^{(96)}_2)^2 \approx - c^2_{(2-6-8)} (e^{\frac{\pi i \tau}{12}} + \cdots),
\label{ME268'}
\end{aligned}
\end{align}
\begin{align}
\begin{aligned}
\lambda_{1}^{(3-5-8)} 
&\approx c^2_{(3-5-8)} 2\sqrt{2} \eta^{(120)}_{1}\eta^{(120)}_{3} \approx 2\sqrt{2} c^2_{(3-5-8)} (e^{\frac{\pi i \tau}{12}} + \cdots), \\
\lambda_{2}^{(3-5-8)} 
&\approx - c^2_{(3-5-8)} 4\sqrt{2} \eta^{(120)}_{6} \eta^{(120)}_{2} \approx - 4\sqrt{2} c^2_{(3-5-8)} (e^{\frac{\pi i \tau}{3}} + \cdots), \\
\lambda_{3}^{(3-5-8)} 
&\approx - c^2_{(3-5-8)} 2\sqrt{2} \eta^{(120)}_{1}\eta^{(120)}_{3} \approx - 2\sqrt{2} c^2_{(3-5-8)} (e^{\frac{\pi i \tau}{12}} + \cdots),
\label{ME358'}
\end{aligned}
\end{align}
\begin{align}
\begin{aligned}
\lambda_{1}^{(4-4-8)} 
&\approx c^2_{(4-4-8)} \left((\eta^{(128)}_{0})^2 + (\eta^{(128)}_{4})^2\right) \approx c^2_{(4-4-8) }(1 + e^{\frac{\pi i \tau}{4}} + \cdots), \\
\lambda_{2}^{(4-4-8)} 
&\approx -c^2_{(4-4-8)} 2\left((\eta^{(128)}_{4})^2 + \eta^{(128)}_{8}\eta^{(128)}_{0}\right) \approx - 2 c^2_{(4-4-8)} (e^{\frac{\pi i \tau}{4}} + e^{\frac{\pi i \tau}{2}} + \cdots), \\
\lambda_{3}^{(4-4-8)} 
&\approx - c^2_{(4-4-8)} \left((\eta^{(128)}_{0})^2 + (\eta^{(128)}_{4})^2\right) \approx - c^2_{(4-4-8) }(1 + e^{\frac{\pi i \tau}{4}} + \cdots).
\label{ME448'}
\end{aligned}
\end{align}

\begin{center}
\begin{figure}[H]
\begin{minipage}{8cm}
\includegraphics[bb=0 0 770 745,width=4.8cm]{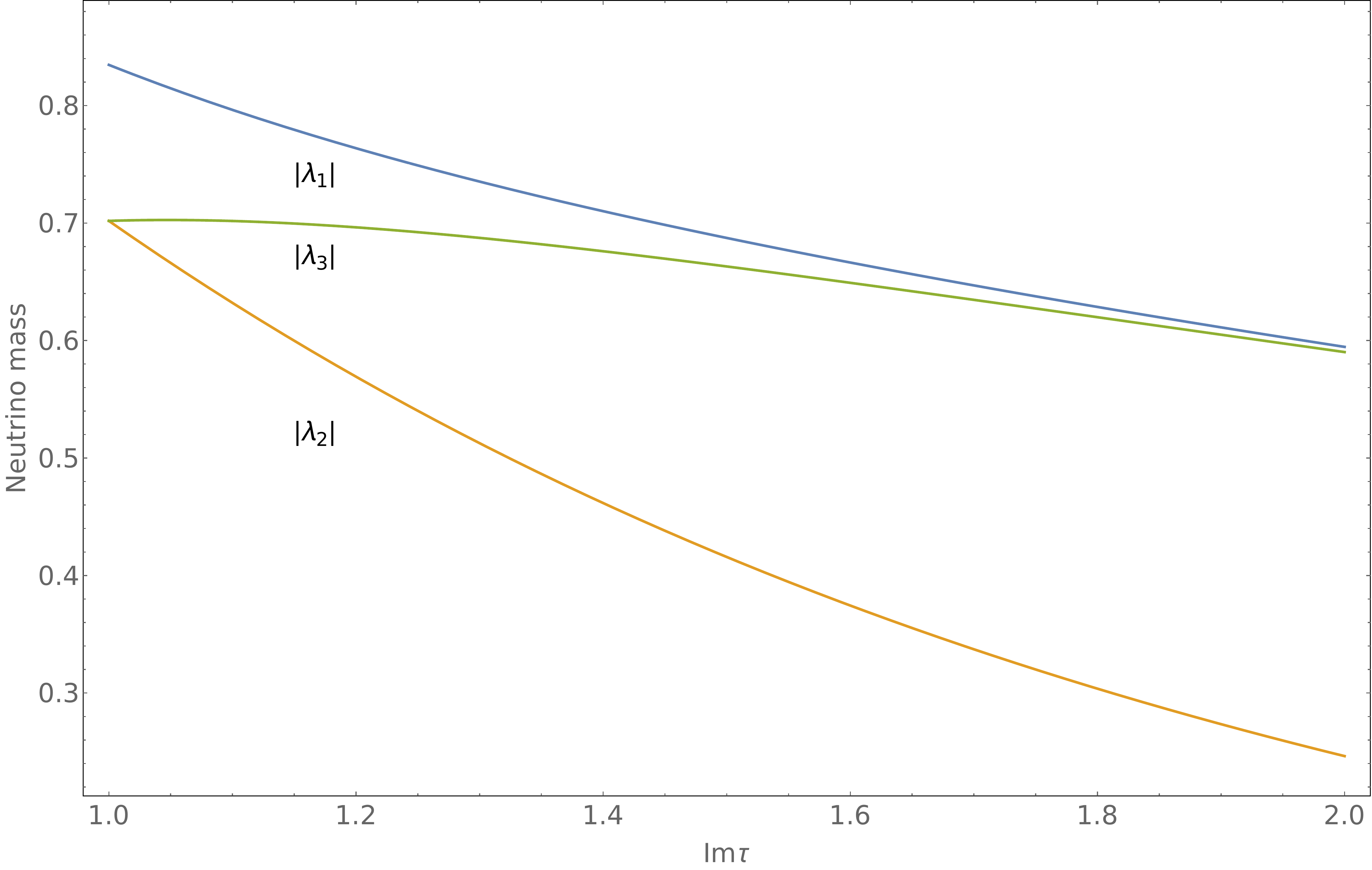}
\caption{The ${\rm Im}\tau$ dependence $(1 \leq {\rm Im}\tau \leq 2)$ of the absolute values of the mass eigenvalues in the 2-6-8 case at ${\rm Re}\tau=0$.}
\label{268Retau0}
\end{minipage}
\hfill
\begin{minipage}{8cm}
\includegraphics[bb=0 0 750 745,width=4.8cm]{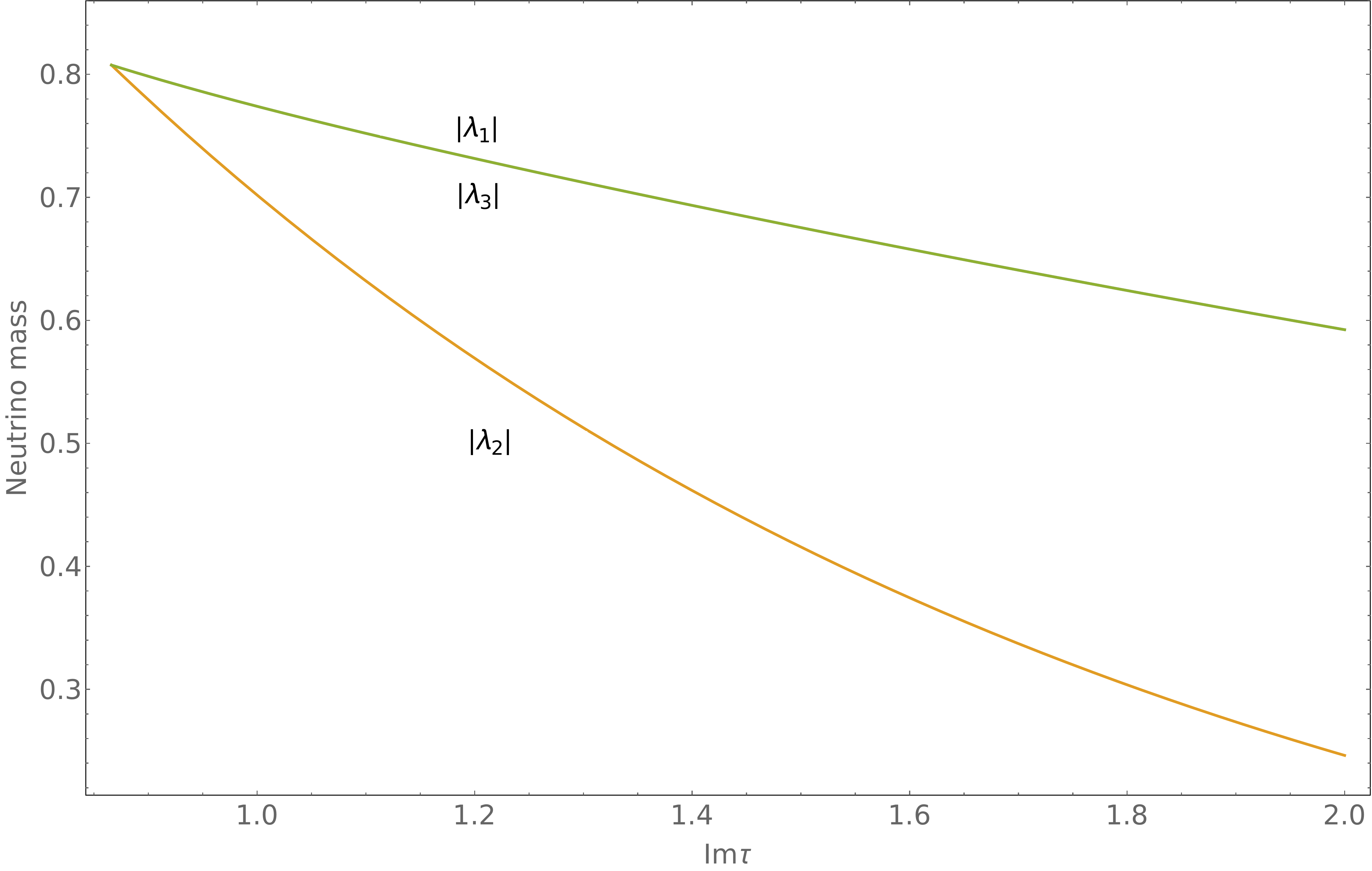}
\caption{The ${\rm Im}\tau$ dependence $(\sqrt{3}/2 \leq {\rm Im}\tau \leq 2)$ of the absolute values of the mass eigenvalues in the 2-6-8 case at ${\rm Re}\tau=1/2$.}
\label{268Retau1/2}
\end{minipage}
\end{figure}
\end{center}

In order to study numerically the behavior of mass eigenvalues $\lambda_{i}^{(M_N = 8)}$ under the change in the modulus $\tau$, let us use the relationship shown by Eq. $({\ref{eq: dominant mass}})$. This allows us to write
\begin{eqnarray}
\lambda_{i}^{(M_N = 8)} \approx e^{-S_{cl}(D_{inst},M_{inst}^{(2-6-8)})}\lambda_{i}^{(2-6-8)} , 
\end{eqnarray}
Thus, we are led to concentrate on the analysis of $\lambda_{i}^{(2-6-8)}$. Hereafter, we omit the overall factor $e^{-S_{cl}(D_{inst},M_{inst}^{(2-6-8)})}$. 
Figures \ref{268Retau0} and \ref{268Retau1/2} show the ${\rm Im}\tau$ dependence $(\sqrt{1-({\rm Re}\tau)^2} \leq {\rm Im}\tau \leq 2)$ of the absolute values of the mass eigenvalues $\lambda_{i}^{(2-6-8)}$ in Eq.~(\ref{ME268}) at ${\rm Re}\tau=0$ and ${\rm Re}\tau=1/2$, respectively.
Here, we set $c_{(2-6-8)}=1$ for simplicity.

There are also four interesting features in Figures \ref{268Retau0} and \ref{268Retau1/2}: $|\lambda_2|=|\lambda_3|$ at $\tau=i$, $|\lambda_1| \simeq |\lambda_3|$ and $|\lambda_2| \rightarrow 0$ at $\tau \rightarrow i \infty$, $|\lambda_1|=|\lambda_3|$ at $\tau=\frac{1}{2}+i{\rm Im}\tau$, and $|\lambda_1|=|\lambda_2|=|\lambda_3|$ at $\tau=\frac{1}{2}+\frac{\sqrt{3}}{2}i$.
As with the 2-2-4 case, they can be expalined by considering that the points $\tau = i$, $i \infty$, $\tau=\frac{1}{2}+i{\rm Im}\tau$, and $\frac{1}{2}+\frac{\sqrt{3}}{2}i$ are invariant under $S$-, $T$-, $T \cdot CP$-, and $ST^{-1}$-transformations, respectively.
These features also appear in $\lambda_{i}^{(3-5-8)}$ and $\lambda_{i}^{(4-4-8)}$ as well as $\lambda_{i}^{(M_N = 8)}$.

\subsection{Neutrino sector with $M_N=5$}
\label{subsec:2-3-5}

Here, we study the mass matrix of the neutrino sector with $M_N=5$.
The  2-3-5 case is the only possible D-brane configuration. 
We take even wavefunctions under $M_{\beta}=2$ and  $M_{\gamma}=3$, for $\beta_i$ and $\gamma_j$ respectively. For the neutrino sector, we take odd wavefunctions under $M_N=5$. 
The $d$ matrices are given by
\begin{align}
\begin{aligned}
d_1 &= c_{(2-3-5)}
\begin{pmatrix}
\sqrt2 \eta^{(30)}_{12} & \eta^{(30)}_2 + \eta^{(30)}_8 \\
\sqrt2 \eta^{(30)}_{3} & \eta^{(30)}_7 + \eta^{(30)}_{13}
\end{pmatrix}, \\
d_2 &= c_{(2-3-5)}
\begin{pmatrix}
\eta^{(30)}_0 & \sqrt2 \eta^{(30)}_{10} \\
\eta^{(30)}_{15} & \sqrt2 \eta^{(30)}_{5}
\end{pmatrix}, \\
d_3 &= c_{(2-3-5)}
\begin{pmatrix}
\sqrt2 \eta^{(30)}_{6} & \eta^{(30)}_4 + \eta^{(30)}_{14} \\
\sqrt2 \eta^{(30)}_{9} & \eta^{(30)}_1 + \eta^{(30)}_{11}
\end{pmatrix}.
\end{aligned}
\end{align}
The mass matrix is then given by
\begin{equation}
\bm{m}^{(2-3-5)}= c^2_{(2-3-5)}
\begin{pmatrix}
X & U & V\\
U & Y &  W\\
V & W & Z
\end{pmatrix},
\end{equation}
where $X,Y,Z,U,V,$and $W$ are defined as
\begin{align}
\begin{aligned}
X &= 2\sqrt{2} \eta^{(30)}_{12}D_7 -  2\sqrt{2} \eta^{(30)}_{3}D_2 \approx - 2\sqrt{2} \eta^{(30)}_{3}\eta^{(30)}_{2} \approx - 2\sqrt{2}(e^{\frac{13\pi i \tau}{30}}+\cdots), \\
Y &= 2\sqrt{2} \eta^{(30)}_0\eta^{(30)}_5  -2\sqrt{2} \eta^{(30)}_{10}\eta^{(30)}_{15} \approx 2\sqrt{2} \eta^{(30)}_0\eta^{(30)}_5 \approx 2\sqrt{2}(e^{\frac{25\pi i \tau}{30}}+\cdots), \\
Z &= 2\sqrt{2} \eta^{(30)}_{6} D_1 - 2\sqrt{2} \eta^{(30)}_{9}D_4 \approx  2\sqrt{2} \eta^{(30)}_{6}\eta^{(30)}_{1} \approx  2\sqrt{2} (e^{\frac{37\pi i \tau}{30}}+\cdots), \\
U &= -\eta^{(30)}_{15} D_2 +2\eta^{(30)}_{5}\eta^{(30)}_{12} + \eta^{(30)}_0D_{7} -2\eta^{(30)}_{3}\eta^{(30)}_{10} \approx \eta^{(30)}_0\eta^{(30)}_7 \approx e^{\frac{49\pi i \tau}{30}}+\cdots, \\
V &= -\sqrt{2}\eta^{(30)}_{3}D_4 + \sqrt{2} \eta^{(30)}_6 D_7 + \sqrt{2} \eta^{(30)}_{12}D_1 - \sqrt{2} \eta^{(30)}_{9}D_2 \approx -\sqrt{2}\eta^{(30)}_{3}\eta^{(30)}_{4} \approx -\sqrt{2}(e^{\frac{25\pi i \tau}{30}}+\cdots), \\
W &= -\eta^{(30)}_{0} D_4 +2\eta^{(30)}_{5}\eta^{(30)}_{6} + \eta^{(30)}_0D_{1} -2\eta^{(30)}_{9}\eta^{(30)}_{10} \approx e^{\frac{\pi i \tau}{30}}+\cdots.
\label{M235}
\end{aligned}
\end{align}
Here, we have defined 
\begin{equation}
D_N = \eta^{(30)}_{N} + \eta^{(30)}_{N+10}.
\end{equation}

\begin{center}
\begin{figure}[H]
\begin{minipage}{8cm}
\includegraphics[bb=0 0 700 615,width=4.6cm]{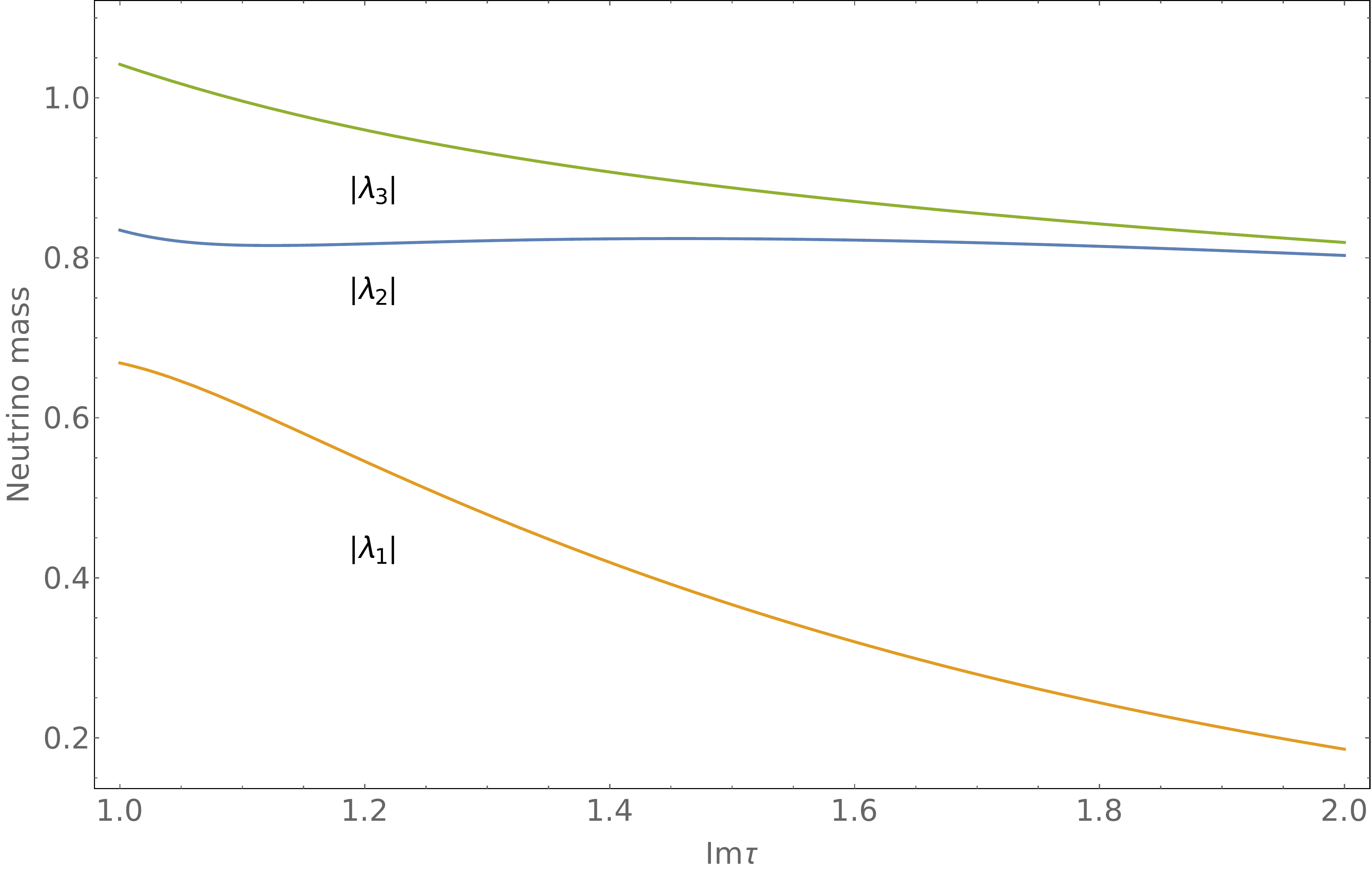}
\caption{The ${\rm Im}\tau$ dependence $(1 \leq {\rm Im}\tau \leq 2)$ of the absolute values of the mass eigenvalues in the 2-3-5 case at ${\rm Re}\tau=0$.}
\label{235Retau0}
\end{minipage}
\hfill
\begin{minipage}{8cm}
\includegraphics[bb=0 0 700 565,width=5.3cm]{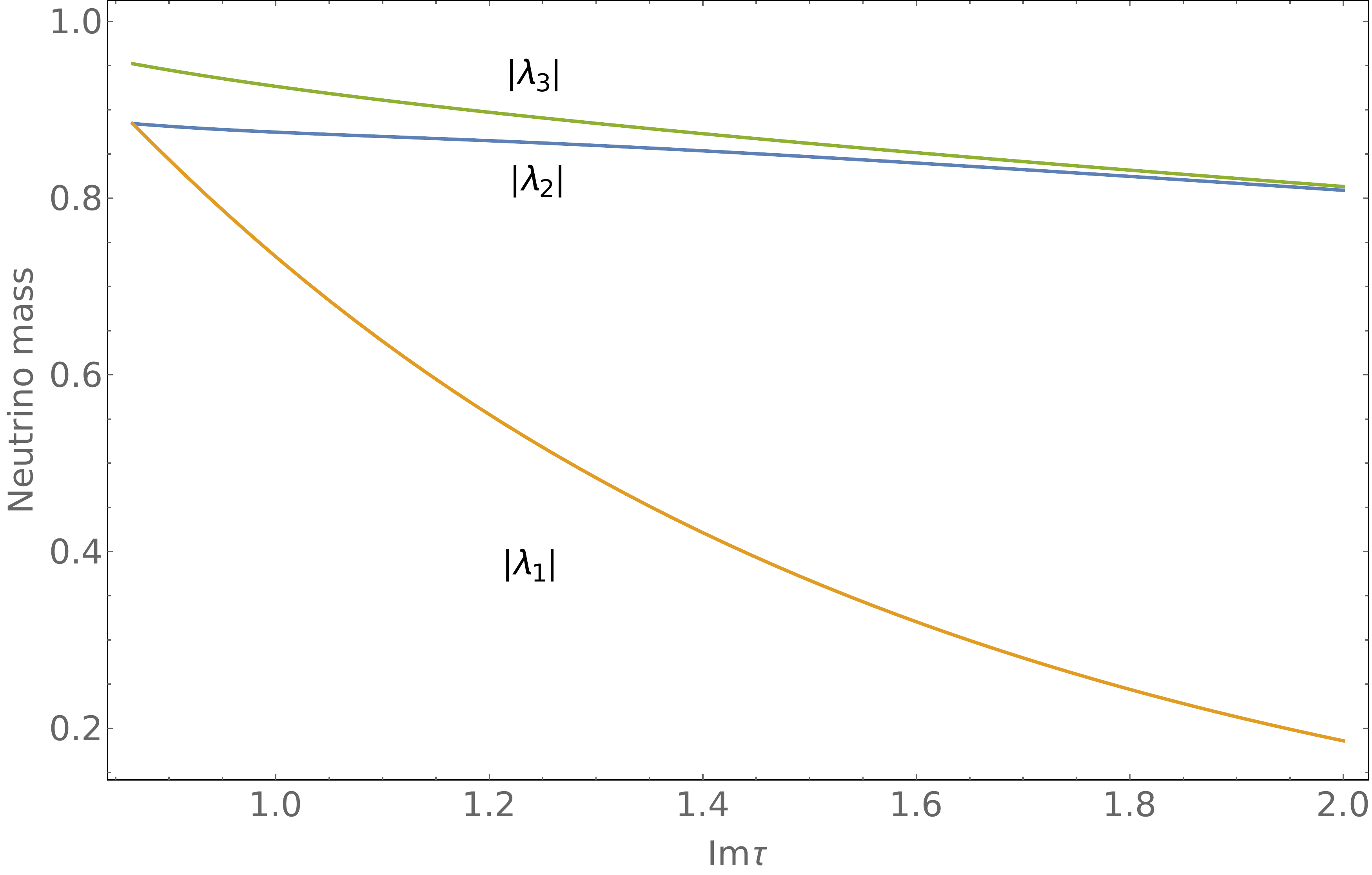}
\caption{The ${\rm Im}\tau$ dependence $(\sqrt{3}/2 \leq {\rm Im}\tau \leq 2)$ of the absolute values of the mass eigenvalues in the 2-3-5 case at ${\rm Re}\tau=1/2$.}
\label{235Retau1/2}
\end{minipage}
\end{figure}
\end{center}

Let us now look at eigenvalues of $\bm{m}^{(2-3-5)}$. 
Unlike the $M_N=4$ and $8$ cases, $\bm{m}^{(2-3-5)}$ is not generally diagonalized by the bimaximal mixing matrix, but the diagonalization depends on the value of the modulus $\tau$.
This may come from that $\mathbb{Z}_2$-shifts are not symmetry  
for the neutrino sector with $M_N=5$ and the SS phase $(0,0)$. That is, the SS phase $(0,0)$ transforms to $(0,1/2)$ and $(1/2,0)$ under $z \rightarrow z+1/2$ and $z \rightarrow z+\tau/2$, respectively\footnote{From Eqs.~(\ref{psiT2}) and (\ref{psiZ2}), we can easily check that
\begin{align}
\psi^{(j+0,0),5}_{T^2/\mathbb{Z}_2^+}(z+\frac{1}{2}) = (-1)^j e^{\frac{5}{2}\pi i\frac{{\rm Im}z}{{\rm Im}\tau}} \psi^{(j+0,\frac{1}{2}),5}_{T^2/\mathbb{Z}_2^+}(z), \quad \psi^{(j+0,0),5}_{T^2/\mathbb{Z}_2^+}(z+\frac{\tau}{2}) = e^{\frac{5}{2}\pi i\frac{{\rm Im}\bar{\tau}z}{{\rm Im}\tau}} \psi^{(2-j+\frac{1}{2},0),5}_{T^2/\mathbb{Z}_2^+}(z). \notag 
\end{align}
}.
Thus, we diagonalize it numerically.
Figures \ref{235Retau0} and \ref{235Retau1/2} show the ${\rm Im}\tau$ dependence $(\sqrt{1-({\rm Re}\tau)^2} \leq {\rm Im}\tau \leq 2)$ of the absolute values of the mass eigenvalues at ${\rm Re}\tau=0$ and ${\rm Re}\tau=1/2$, respectively.
Here, we set $c_{(2-3-5)}=1$ for simplicity.

There are also several interesting features.
First, since the point $\tau=i$ is invariant under $S$-transformation, the mass eigenvalues at $\tau=i$ must be invariant under $S$-transformation.
Then, the mass eigenstates must be also invariant under $S$-transformation.
We note that $S$-transformation at $\tau=i$ induces $\mathbb{Z}_4$-twist for the coordinate of $T^2$,~i.e. $z \rightarrow iz$.
Hence, the mass eigenstates must be $\mathbb{Z}_4$-twisted eigenstates.
The diagonalizing matrix $P$ at $\tau=i$ satisfying $P^{\mathrm T} \bm{m}^{(2-3-5)} P = {\rm diag}(\lambda_1, \lambda_2, \lambda_3)$ corresponds to the unitary transformation from the $\mathbb{Z}_2$-twisted basis to the $\mathbb{Z}_4$-twisted basis,~i.e. $\psi_{T^2/\mathbb{Z}_4} = P^{\mathrm T} \psi_{T^2/\mathbb{Z}_2}$, given as
\begin{equation}
P = 
\begin{pmatrix}
\sqrt{\frac{23-3\sqrt{5}}{30-2\sqrt{5}}} & -\frac{1}{\sqrt{5-\sqrt{5}}} & 0 \\
\frac{2}{\sqrt{30-2\sqrt{5}}} & \sqrt{\frac{3-\sqrt{5}}{5-\sqrt{5}}} & \sqrt{\frac{3+\sqrt{5}}{7+\sqrt{5}}} \\
-\sqrt{\frac{3+\sqrt{5}}{30-2\sqrt{5}}} & -\frac{1}{\sqrt{5-\sqrt{5}}} & \frac{2}{\sqrt{7+\sqrt{5}}}
\end{pmatrix}.
\label{Z4Z2}
\end{equation}
This is nearly the tri-bimaximal mixing matrix, i.e.
\begin{align}
|P_{ij}|^2 \approx 
\begin{pmatrix}
0.64& 0.36 & 0 \\
0.16& 0.28&  0.57\\
0.21& 0.36 & 0.43
\end{pmatrix}.
\end{align}
Its derivation is shown in Appendix~\ref{app:Z4}.
Actually, this is consistent with the neumerical analysis.

Second, we can find $|\lambda_2| \simeq |\lambda_3|$ and $|\lambda_1| \rightarrow 0$ at $\tau \rightarrow i \infty$, $\frac{1}{2} + i \infty$. Moreover, the mass matrices $\bm{m}^{(2-3-5)}$ are diagonalized by almost the bimaximal mixing matrix .
They can be explained by considering $T$-invariance.
Note that since the neutrinos with $M_N=5$ and the SS phase $(0,0)$ transform into ones with the SS phase $(0,1/2)$ under $T$-transformation\footnote{See in detail Ref.~\cite{Kikuchi:2021ogn}.}, we cannot consider $T$-transformation for those neutrinos in general.
However, in the limit ${\rm Im}\tau \rightarrow \infty$, the SS phases can be negligible\footnote{In this limit, the shifts $z \rightarrow z+1$ and $z \rightarrow z+2$ cannot be distinguished, which means the difference of the SS phase $\alpha_1$ can be negligible. Similarly, the shifts $z \rightarrow z+\frac{1}{2}$ and $z \rightarrow z+1$ cannot be distinguished, which means the difference of the SS phase $\alpha_{\tau}$ can be negligible.}.
In the limit ${\rm Im}\tau \rightarrow \infty$, 
there appears approximately $T$-transformation as well as $\mathbb{Z}_2$-shifts for the neutrinos.
Then, the mass eigenvalues at $\tau \rightarrow i \infty$, $\frac{1}{2}+i \infty$ must be almost invariant under $T$-transformation as well as $\mathbb{Z}_2$-shifts.
Thus, as with the $M_N=4$ and $8$ cases, the mass eigenstates at $\tau \rightarrow i \infty$, $\frac{1}{2} + i \infty$ become $\mathbb{Z}_2$-twisted and almost $\mathbb{Z}_2$-shifted eigenstates, which means the mass matrix is diagonalized by almost the bimaximal mixing matrix.
Moreover, since the eigenstates of $T$-transformation are not the $\mathbb{Z}_2$-twisted and (almost) $\mathbb{Z}_2$-shifted basis but the $\mathbb{Z}_2$-twisted basis, two eigenvalues for two mass eigenstates transformed by almost the bimaximal mixing matrix become same,~i.e. $|\lambda_2| \simeq |\lambda_3|$.
Actually, from Eq.~(\ref{M235}), the mass matrix $\bm{m}^{(2-3-5)}$ at $\tau \rightarrow i \infty$, $\frac{1}{2} + i \infty$ is estimated as
\begin{equation}
\bm{m}^{(2-3-5)} \approx e^{\frac{\pi i \tau}{30}}
\begin{pmatrix}
-2\sqrt{2} e^{\frac{2\pi i \tau}{5}} & 0 & 0\\
0 & 0 & 1\\
0 & 1 & 0
\end{pmatrix}.
\end{equation}
Then, it is diagonalized by the bimaximal mixing matrix and the eigenvalues satisfy $\lambda_2 \simeq - \lambda_3$ and $\lambda_1 \rightarrow 0$ at $\tau \rightarrow i \infty$, $\frac{1}{2} + i \infty$.

Third, we can find $|\lambda_1|=|\lambda_2|$ at $\tau=\frac{1}{2}+\frac{\sqrt{3}}{2}i$.
It can be explained by considering that the point $\tau=\frac{1}{2}+\frac{\sqrt{3}}{2}i$ is invariant under $ST^{-1}$- and $T \cdot CP$-transformations.
However, unlike $M_N=4$ and $8$ cases, $|\lambda_3|$ is not the same as $|\lambda_2|$. In addition, the diagonalizing matrix departs from the bimaximal mixing matrix slightly, though it can be still approximated by the bimaximal mixing matirx.
This may be because the difference of the SS phases by $T$-transformation as well as $\mathbb{Z}_2$-shifts cannot be ignored though we can consider the simultaneous transformation of $T$-transformation and $\mathbb{Z}_2$-shifts.
Then, the point $\tau=\frac{1}{2}+\frac{\sqrt{3}}{2}i$ is also invariant under $S \cdot CP$-transformation.
Actually, on the mass eigenstates, $|\lambda_1|$ and $|\lambda_2|$ are exchanged under $S \cdot CP$-transformation, while $|\lambda_3|$ is invariant.
It means that $|\lambda_1|=|\lambda_2|$ at $\tau=\frac{1}{2}+\frac{\sqrt{3}}{2}i$ is required.

\subsection{Neutrino sector with $M_N=7$}

Here, we study mass matrix for the neutrino sector with $M_N=7$.
For this neutrino sector, 
the mass terms can be induced by the 2-5-7 case and the 3-4-7 case.

\subsubsection{2-5-7 Case}
In the 2-5-7 case, we take even wavefunctions under $M_{\beta}=2$ and odd wavefunctions under $M_{\gamma}=5$, for $\beta_i$ and $\gamma_j$ respectively. For the neutrino sector, we take odd wavefunctions under $M_N=7$. 
The $d$ matrices are given by
\begin{eqnarray}
d_1 =  c_{(2-5-7)}
\begin{pmatrix}
-E_{18} & E_4 \\
E_{-3} & -E_{11}
\end{pmatrix}
, \
d_2 = c_{(2-5-7)}
\begin{pmatrix}
D_2 & -D_{16} \\
-D_{23} & D_{9} \\
\end{pmatrix}
, \
d_3 = c_{(2-5-7)}
\begin{pmatrix}
F_{-8} & -F_{6} \\
-F_{13} & F_{-1} 
\end{pmatrix}.
\end{eqnarray}
Here, $D_N, E_N$, and $F_N$ are defined as
\begin{align}
\begin{aligned}
D_N &= \eta^{(70)}_N - \eta^{(70)}_{N+10}, \\
E_N &= \eta^{(70)}_N - \eta^{(70)}_{N+20}, \\
F_N &= \eta^{(70)}_N - \eta^{(70)}_{N+30}.
\end{aligned}
\end{align}
The mass matrix is then given by
\begin{equation}
\bm{m}^{(2-5-7)}=c^2_{(2-5-7)}
\begin{pmatrix}
X  & U &   V\\
U  & Y & W\\
V & W & Z
\end{pmatrix},
\end{equation}
where $X,Y,Z,U,V,$ and $W$ are defined as
\begin{align}
\begin{aligned}
X &=  2 E_{18}E_{11} - 2E_{-3}E_{4} \approx -2\eta^{(70)}_3\eta^{(70)}_4 \approx -2( e^{\frac{5 \pi i \tau}{14}} + \cdots),\\
Y &=  -2D_{23}D_{16} +2D_2D_9 \approx 2\eta^{(70)}_2\eta^{(70)}_9 \approx 2(e^{\frac{17 \pi i \tau}{14}} + \cdots),  \\
Z &=  2 F_{-8}F_{-1} - 2 F_{13} F_{6} \approx 2\eta^{(70)}_8\eta^{(70)}_1 \approx 2(e^{\frac{13 \pi i \tau}{14}} + \cdots),  \\
U &= D_{23}E_{4} - D_9E_{18} -D_2 E_{11} +D_{16} E_{-3} \approx -\eta^{(70)}_2\eta^{(70)}_{11} \approx -( e^{\frac{25 \pi i \tau}{14}} + \cdots), \\
V &= E_{-3}F_6 - E_{11}F_{-8} - E_{18}F_{-1} + E_{4}F_{13} \approx \eta^{(70)}_3\eta^{(70)}_6 \approx e^{\frac{9 \pi i \tau}{14}} + \cdots, \\
W &= -D_{23}F_6 + D_9F_{-8} + D_2F_{-1} - D_{16}F_{13} \approx \eta^{(70)}_2\eta^{(70)}_1 \approx e^{\frac{\pi i \tau}{14}} + \cdots.
\end{aligned}
\end{align}

\subsubsection{3-4-7 Case}
In 3-4-7 case, there are three possible variations with different SS phases, $(1/2,0), (0,1/2)$, and $(1/2,1/2)$.
The sum of three mass matrices in equal ratio is given by
\begin{equation}
\label{eq: 4-4-8}
\begin{split}
\bm{m}^{(3-4-7)}=&\bm{m}^{(3-4-7)(1/2,0)} + \bm{m}^{(3-4-7)(0,1/2)} + \bm{m}^{(3-4-7)(1/2,1/2)} 
\\=&  c_{(3-4-7)}^2
\begin{pmatrix}   
X^{(3-4-7)} & U^{(3-4-7)} & V^{(3-4-7)} \\
U^{(3-4-7)}  & Y^{(3-4-7)}  & W^{(3-4-7)} \\
V^{(3-4-7)} & W^{(3-4-7)}  & Z^{(3-4-7)} 
\end{pmatrix},
\end{split}
\end{equation}
where 
\begin{align}
\begin{aligned}
X^{(3-4-7)} &= 2( S_{(-9,-61)}+ R_{(30,44)} ) \approx -2\sqrt{2}\eta^{(336)}_{5}\eta^{(336)}_{9} \approx -2\sqrt{2} e^{\frac{53 \pi i \tau}{168}}, \\
Y^{(3-4-7)} &=  2 ( S_{(-57,-13)} + R_{(6,20)}) \approx 2\sqrt{2}\eta^{(336)}_{13}\eta^{(336)}_{15} \approx 2\sqrt{2} e^{\frac{197 \pi i \tau}{168}},\\
Z^{(3-4-7)} &= 2( S_{(-81,11)}+ R_{(-18,-4)}) \approx 2\sqrt{2}\eta^{(336)}_{3}\eta^{(336)}_{17} \approx 2\sqrt{2} e^{\frac{149 \pi i \tau}{168}},\\
U^{(3-4-7)} &= S_{(-9,-13)} + S_{(15,-37)} + R_{(6,44)} + R_{(30,20)} \approx -\sqrt{2}\eta^{(336)}_{15}\eta^{(336)}_{19} \approx -\sqrt{2} e^{\frac{293 \pi i \tau}{168}}, \\
V^{(3-4-7)} &= S_{(-33,11)} + S_{(39,-61)} + R_{(30,-4)} + R_{(-18,44)} \approx \sqrt{2}\eta^{(336)}_{9}\eta^{(336)}_{11} \approx \sqrt{2} e^{\frac{101 \pi i \tau}{168}}, \\
W^{(3-4-7)} &= S_{(39,-13)} + S_{(15,11)} + R_{(6,-4)} + R_{(-18,20)} \approx \sqrt{2}\eta^{(336)}_{3}\eta^{(336)}_{1} \approx \sqrt{2} e^{\frac{5 \pi i \tau}{168}}.
\end{aligned}
\end{align}
Here we defined
\begin{align}
\begin{aligned}
S_{(M,N)} &= \frac{1}{\sqrt2}(Q_{(M,N)} - P_{(M,N)}), \\
P_{(M,N)} &= B_M (B_N + B_{N+56}) - B_{14-N}(B_{14-M} + B_{(14-M)+56}), \\
Q_{(M,N)} &= E_M ( D_N - D_{N+56}) + D_{14-N} (E_{14-M} - E_{(14-M)+56}), \\
R_{(M,N)} &= G_M F_N - A_{M-42} (F_{N-42} + F_{(N-42)+84}), \\
A_N &= \eta^{(336)}_N - \eta^{(336)}_{N+168}, \\
B_N &= (\eta^{(336)}_N + \eta^{(336)}_{N+168}) - (\eta^{(336)}_{N+42} + \eta^{(336)}_{(N+42)+168}), \\
D_N &= A_N - A_{N+42}, \\
E_N &= A_N + A_{N+42}, \\
F_N &= {A}_N - {A}_{N+56}, \\
G_N &= {A}_N - {A}_{N+84}. 
\end{aligned}
\end{align}

\subsubsection{Full mass matrix and and mass eigenvalues}
\begin{center}
\begin{figure}[H]
\begin{minipage}{8cm}
\includegraphics[bb=0 0 760 665,width=5cm]{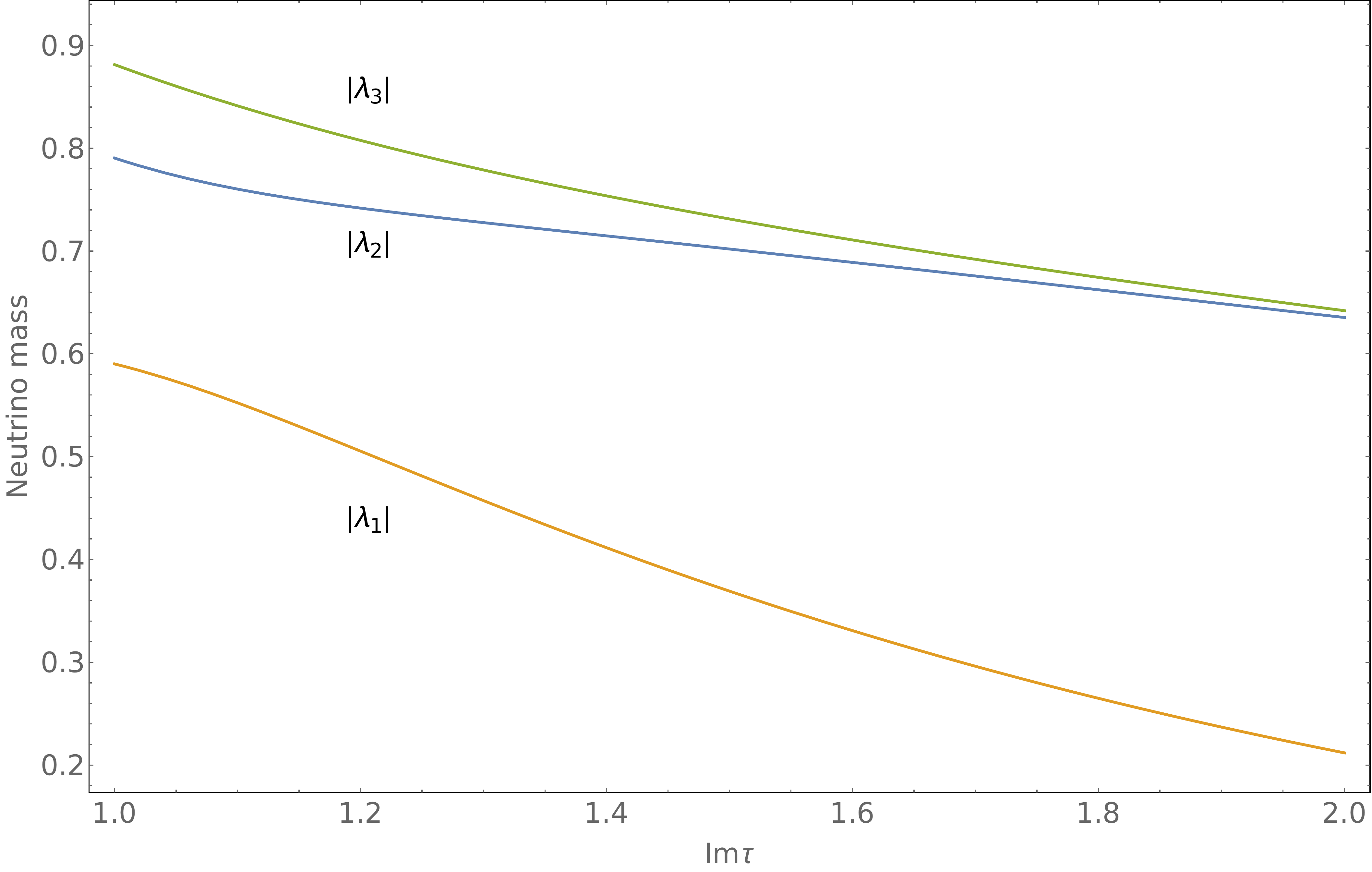}
\caption{The ${\rm Im}\tau$ dependence $(1 \leq {\rm Im}\tau \leq 2)$ of the absolute values of the mass eigenvalues in the 2-5-7 case at ${\rm Re}\tau=0$.}
\label{257Retau0}
\end{minipage}
\hfill
\begin{minipage}{8cm}
\includegraphics[bb=0 0 730 615,width=5.5cm]{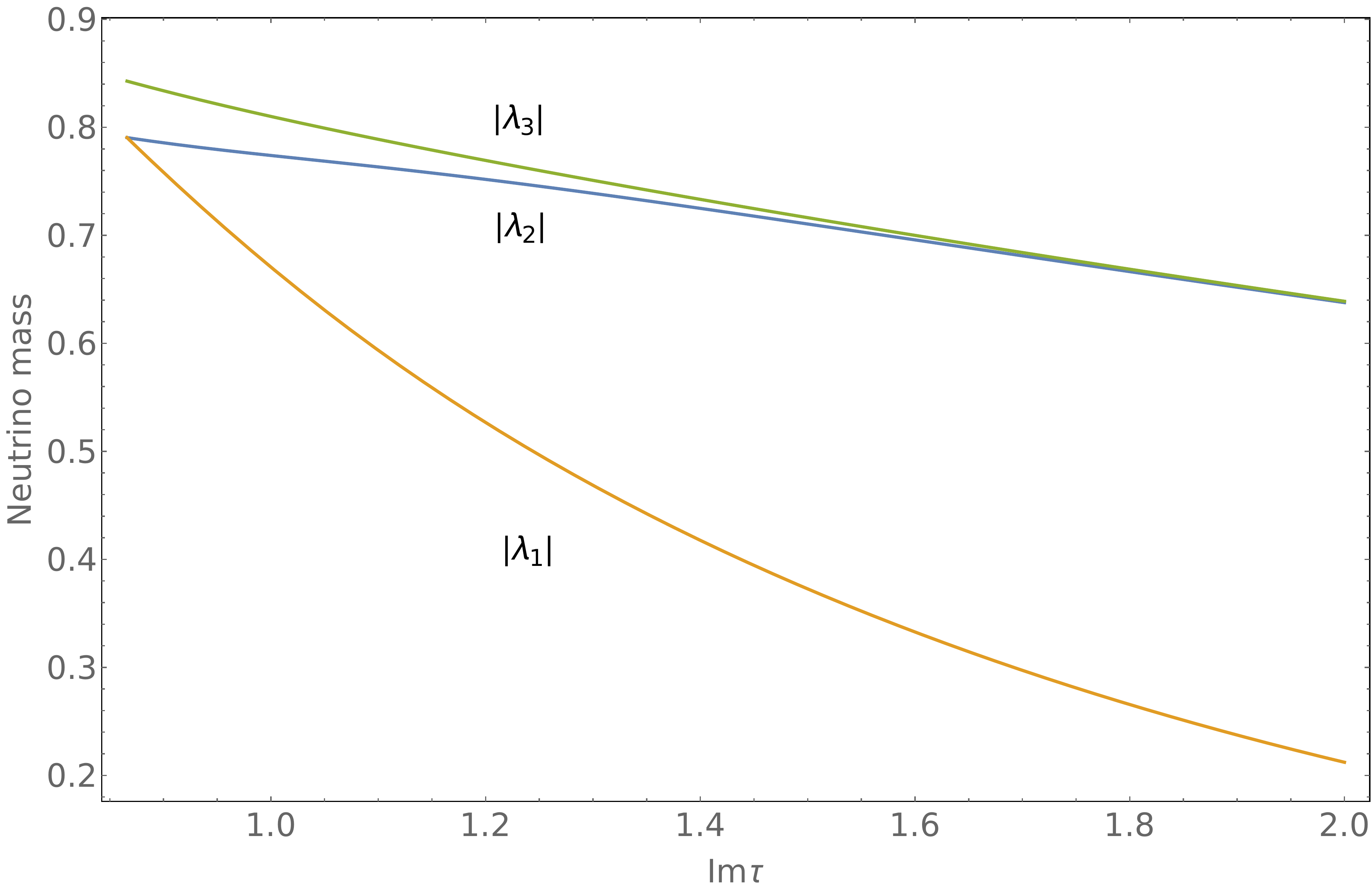}
\caption{The ${\rm Im}\tau$ dependence $(\sqrt{3}/2 \leq {\rm Im}\tau \leq 2)$ of the absolute values of the mass eigenvalues in the 2-5-7 case at ${\rm Re}\tau=1/2$.}
\label{257Retau1/2}
\end{minipage}
\end{figure}
\end{center}

Finally, let us look at mass eigenvalues of the full mass matrix $\bm{M}^{(M_N=7)}$.
Unlike the $M_N=8$ case, diagonalization of  $\bm{M}^{(M_N=7)}$ depends on the factors $e^{-S_{cl}(D_{inst},M_{inst})}$ as well as the modulus $\tau$ in general.
Here, we evaluate $\bm{M}^{(M_N=7)}$ with $\bm{m}^{(2-5-7)}$ as Eq.~(\ref{evM7}).
Then, we diagonalize $\bm{m}^{(2-5-7)}$ numerically, where we omit the overall factor $e^{-S_{cl}(D_{inst},M_{inst})}$.
Figures \ref{257Retau0} and \ref{257Retau1/2} show the ${\rm Im}\tau$ dependence $(\sqrt{1-({\rm Re}\tau)^2} \leq {\rm Im}\tau \leq 2)$ of the absolute values of the mass eigenvalues at ${\rm Re}\tau=0$ and ${\rm Re}\tau=1/2$, respectively.
Here, we set $c_{(2-5-7)}=1$ for simplicity.
Similarly, there are several features and they can be expalined as with the $M_N=5$ case.
These features also appear in $\bm{m}^{(3-4-7)}$ as well as $\bm{M}^{(M_N=7)}$.
In particular, it is interesting that in the large ${\rm Im}\tau$ limit, both matrices $\bm{m}^{(2-5-7)}$ and $\bm{m}^{(3-4-7)}$
can be approximated by
\begin{align}
\bm{m}^{(2-5-7)}\approx a_{(2-5-7)}\bm{m}^{(7)},\qquad \bm{m}^{(3-4-7)}\approx a_{(3-4-7)}\bm{m}^{(7)},
\end{align}
where
\begin{align}
\bm{m}^{(7)}=
\begin{pmatrix}
-2q_7 & -q_7^6&  q_7^2 \\
-q_7^6& 2q_7^4 &1 \\
q_7^2 & 1 & 2q_7^3
\end{pmatrix},\qquad q_7=e^{2 \pi i \tau/7}.
\end{align}
It seems that there appears the $\mathbb{Z}_{14}$ symmetry originated from the $T$-transformation 
in the matrix $\bm{m}^{(7)}$.
In addition, overall factors $a_{(2-5-7)}$ and $a_{(3-4-7)}$ are written by 
\begin{align}
a_{(2-5-7)} = e^{\pi i \tau/14}, \qquad a_{(3-4-7)} = \sqrt{2} e^{5 \pi i \tau/168}.
\end{align}
That leads to different phases under the $T$-transformation \footnote{
The anomaly would be relevant to these structures. 
The study on modular symmetry anomaly is beyond our scope.
We would study elsewhere.}.

\section{Conclusion}
\label{sec:conclusion}

We have studied Majorana neutrino masses induced by 
D-brane instanton effects in magnetized orbifold models.
We have systematically studied 
the D-brane configutions, where neutrino masses can be induced.
Three and four generations are favorable in order to 
generate Majorana neutrino masses by D-brane instanton effects.
Also we have computed explicit patterns of neutrino mass matrices.
These matrices have specific features.
Our basis is the $T^2/\mathbb{Z}_2$ orbifold basis.
The diagonalizing matrices of neutrino mass matrices 
are the bimaximal mixing matrix in the case with even magnetic fluxes, 
independently of the modulus value $\tau$.
On the other hand, for odd magnetic fluxes, 
the diagonalizing matrices correspond nearly to  the tri-bimaximal mixing matrix 
near $\tau = i$, while they becomes bimaximal one 
for larger ${\rm Im}\tau$ \footnote{The point $\tau = i$ may be favorable from the viewpoint of 
the modulus stabilization \cite{Ishiguro:2020tmo}.}.

For even magnetic fluxes, 
neutrino masses are modular forms of the
weight 1 on $T^2/\mathbb{Z}_2$, and they have symmetries such as 
$S_4'$ and  $\Delta '(96)\times \mathbb{Z}_3$.
These modular form structures of  Majorana neutrino masses can provide us with the UV-completion for recent bottom-up approach 
to construct modular flavor symmetric models \cite{Feruglio:2017spp}.
We can extend our analysis on the $T^4/\mathbb{Z}_2$ orbifold with two moduli, $\tau_1$ and $\tau_2$.
When we identify $\tau=\tau_1=\tau_2$ like in Ref.~\cite{Kikuchi:2020nxn}, 
Majorana neutrino masses would correspond to modular forms of weight 2 with symmetries, $\Gamma_N$.

The patterns of Majorana neutrino mass matrices are quite interesting.
For example, we can realize almost the tri-bimaximal mixing matrix as the diagonalizing matrix 
in our $T^2/\mathbb{Z}_2$ orbifold basis.
However, this diagonalizing matrix is not physical observables.
We have to examine the charged lepton mass matrix and Dirac neutrino mass matrix.
Then, we can discuss the mixing angles in the lepton sector, i.e. the PMNS matrix.
Three-generation models have been studied in Refs.~\cite{Abe:2008sx,Abe:2015yva,Hoshiya:2020hki}.
It is very interesting to  combine those analyses with our results in this paper to analyze 
light neutrino masses and the mixing angles in the lepton sector.
We would study it elsewhere.

It is also important to extend our analysis to other $T^2/\mathbb{Z}_N$ orbifolds \cite{Abe:2013bca,Abe:2014noa} 
as well as resolved orbifolds \cite{Kobayashi:2019fma}.
That is beyond our scope.

\vspace{1.5 cm}
\noindent
{\large\bf Acknowledgement}\\

T. K. was supported in part by MEXT KAKENHI Grant Number JP19H04605. 
H. U. was supported by Grant-in-Aid for JSPS Research Fellows No. 20J20388.


\appendix

\section{Modular symmetry}
\label{app:moduar}

Here, we review the modular symmetry on $T^2$ and modular forms~\cite{Gunning:1962,Schoeneberg:1974,Koblitz:1984,Bruinier:2008}.
The modular transformation for the modulus $\tau$ as well as the coordinate $z$ of $T^2$ are defined as
\begin{align}
\gamma: \tau \rightarrow \gamma(\tau)=\frac{a\tau+b}{c\tau+d}, \quad \gamma: z \rightarrow \gamma(z)=\frac{z}{c\tau+d}, \quad \gamma=
\begin{pmatrix}
a & b \\
c & d
\end{pmatrix}
\in SL(2,\mathbb{Z}) \equiv \Gamma.
\end{align}
They are generated by the following $S$ and $T$ transformation:
\begin{align}
\begin{aligned}
&S: \tau \rightarrow S(\tau)=-\frac{1}{\tau}, \quad S: z \rightarrow S(z)=-\frac{z}{\tau}, \quad S=
\begin{pmatrix}
0 & 1 \\
-1 & 0
\end{pmatrix}
\in \Gamma, \\
&T: \tau \rightarrow T(\tau)=\tau+1, \quad T: z \rightarrow T(z)=z, \quad T=
\begin{pmatrix}
1 & 1 \\
0 & 1
\end{pmatrix}
\in \Gamma .
\end{aligned}
\end{align}
They satisfy the following relations,
\begin{align}
S^4=(ST)^3=\mathbb{I}.
\end{align}
In particular, $S^2=-\mathbb{I}$ transformation for the modulus $\tau$ is identified with the identity $\mathbb{I}$.
In this sense, $\bar{\Gamma} \equiv \Gamma/\{\pm\mathbb{I}\}$ is called (inhomogeneous) modular group.
Note that $CP$ transformation for the modulus $\tau$~\cite{Baur:2019kwi,Novichkov:2019sqv} as well as the coordinate $z$ are also defined as
\begin{align}
CP: \tau \rightarrow CP(\tau) = -\bar{\tau}, \quad CP: z \rightarrow CP(z) = -\bar{z}.
\end{align}
Then, the fundamental region, $F$, of the modulus $\tau$ becomes
\begin{align}
F = \left\{ \tau \in \mathbb{C} \left| -\frac{1}{2} \leq {\rm Re}\tau \leq \frac{1}{2}, \ \sqrt{1-({\rm Re}\tau)^2} \leq {\rm Im}\tau \right. \right\}.
\end{align}
Table~\ref{gammatau} shows the specific points of modulus $\tau$ and the operators $\gamma \in \bar{\Gamma}$ satisfying $\gamma(\tau) = \tau$.
\begin{table}[H]
\centering
\begin{tabular}{|c|c|} \hline
$\tau$ & $\gamma$ \\ \hline
$i$ & $S$ \\ \hline
$\frac{1}{2}+\frac{\sqrt{3}}{2}i$ ($-\frac{1}{2}+\frac{\sqrt{3}}{2}i$) & $ST^{-1}$, $T \cdot CP$, $S \cdot CP$ ($ST$, $T^{-1} \cdot CP$, $S \cdot CP$) \\ \hline
$\frac{1}{2}+i{\rm Im}\tau$ ($-\frac{1}{2}+i{\rm Im}\tau$) & $T \cdot CP$ ($T^{-1} \cdot CP$) \\ \hline
$({\rm Re}\tau)+i \infty$ & $T$ \\ \hline
\end{tabular}
\caption{The modulus $\tau$ and the operators $\gamma$ satisfying $\gamma(\tau)=\tau$.}
\label{gammatau}
\end{table}

Now, let us review the modular forms.
First, we introduce the principal congruence subgroup of level $N$ defined as
\begin{align}
  \Gamma (N) \equiv \left\{ h=
  \begin{pmatrix}
    a' & b' \\
    c' & d'
  \end{pmatrix}
  \in \Gamma \left|
  \begin{pmatrix}
    a' & b' \\
    c' & d'
  \end{pmatrix}
  \equiv
  \begin{pmatrix}
    1 & 0 \\
    0 & 1
  \end{pmatrix}
  \right.
  ({\rm mod}\ N)
  \right\}.
\end{align}
The modular forms, $f(\tau)$, of the integral weight $k$ for $\Gamma(N)$ are the holomorphic functions of  $\tau$ transforming under the modular transformation as
\begin{align}
\begin{aligned}
  &f(\gamma(\tau)) = J_k(\gamma,\tau) \rho(\gamma) f(\tau), \quad J_k(\gamma,\tau) = (c\tau + d)^k, \quad
\gamma = 
  \begin{pmatrix}
    a & b \\
    c & d
  \end{pmatrix}
  \in \Gamma, \\
  &f(h(\tau)) = J_k(h,\tau) f(\tau), \quad J_k(h,\tau) = (c'\tau + d')^k, \quad \rho(h)=\mathbb{I}, \quad
h = 
  \begin{pmatrix}
    a' & b' \\
    c' & d'
  \end{pmatrix}
  \in \Gamma(N),
\end{aligned}
\end{align}
where $J_k(\gamma,\tau)$ is the automorphy factor and $\rho$ is the unitary representation of the quotient group $\Gamma'_N \equiv \Gamma/\Gamma(N)$, satisfying the following relations
\begin{align}
 \rho(Z) = \rho(S)^2 = (-1)^k\mathbb{I}, \ \rho(Z)^2 = \rho(S)^4 = [\rho(S)\rho(T)]^3 = \mathbb{I}, \ \rho(T)^N = \mathbb{I}, \ \rho(Z)\rho(T) = \rho(T)\rho(Z).
\end{align}
In particular, in the case of even weight $k$, $\rho$ is also the unitary representation of the finite modular subgroup $\Gamma_N \equiv \bar{\Gamma}/\bar{\Gamma}(N)$, where $\bar{\Gamma}(N) \equiv \Gamma(N)/\{\pm\mathbb{I}\}$ for $N=1,2$ and $\bar{\Gamma}(N) \equiv \Gamma(N)$ for $N>2$.
Moreover, it is well known that $\Gamma_2 \simeq S_3$, $\Gamma_3 \simeq A_4$, $\Gamma_4 \simeq S_4$, $\Gamma_5 \simeq A_5$~\cite{deAdelhartToorop:2011re}, and also $\Gamma'_N$ becomes the double covering group of $\Gamma_N$~\cite{Liu:2019khw}.

\section{$\mathbb{Z}_4$-twisted basis}
\label{app:Z4}
  
Here, we derive the basis transformation in Eq.~(\ref{Z4Z2}) from the $\mathbb{Z}_2$-twisted basis into the $\mathbb{Z}_4$-twisted basis.
First, when we consider the $T^2/\mathbb{Z}_4$-twisted orbifold, the modulus must be $\tau=i$.
The $\mathbb{Z}_4$-twisted eigenstates with eigenvalues $p=\pm 1,\pm i$ and SS phase $(0,0)$~\cite{Kobayashi:2017dyu,Abe:2013bca,Abe:2014noa} can be constructed by
\begin{align}
\psi_{T^2/\mathbb{Z}_4^p}(z)
&= {\cal N} \left( \psi^{(j,0),|M|}_{T^2}(z) + p^{-1} \psi^{(j,0),|M|}_{T^2}(iz) + p^{-2} \psi^{(j,0),|M|}_{T^2}(-z) + p^{-3} \psi^{(j,0),|M|}_{T^2}(-iz) \right) \notag \\
&= \sum_{k=0}^{|M|-1} {\cal N} \left( \left( \delta_{j,k} + p^{-2} \delta_{|M|-j,k} \right) + \frac{p^{-1}}{\sqrt{|M|}} \left( e^{\frac{2\pi ijk}{|M|}} + p^{-2} e^{-\frac{2\pi ijk}{|M|}} \right) \right) \psi^{(k,0),|M|}_{T^2}(z),
\end{align}
where ${\cal N}$ denotes the normalization factor.
Here, we use the $S$-transformation for wavefunctions on $T^2$,
\begin{align}
\psi^{(j,0),|M|}_{T^2}(-\frac{z}{\tau}, -\frac{1}{\tau}) &= (-\tau)^{1/2} \sum_{k=0}^{|M|-1} \frac{{e^{\frac{\pi i}{4}}}}{\sqrt{M}} e^{\frac{2\pi ijk}{|M|}} \psi^{(k,0),|M|}_{T^2}(z,\tau),
\end{align}
at $\tau=i$ since the $\mathbb{Z}_4$-twist, $z \rightarrow iz$, can be induced by $S$-transformation at $\tau=i$.
Then, the number of zero-modes is given in Table~\ref{Z4}.
\begin{table}[H]
\centering
\begin{tabular}{|c|c|c|c|c|} \hline
$|M|$ & $4n$ & $4n+1$ & $4n+2$ & $4n+3$ \\ \hline
$p=1$ & $n+1$ & $n+1$ & $n+1$ & $n+1$ \\ \hline
$p=-1$ & $n$ & $n$ & $n+1$ & $n+1$ \\ \hline
$p=i$ & $n$ & $n$ & $n$ & $n+1$ \\ \hline
$p=-i$ & $n-1$ & $n$ & $n$ & $n$ \\ \hline
\end{tabular}
\caption{The number of zero-modes of $\mathbb{Z}_4$-twisted eigenmodes with SS phase $(0,0)$.}
\label{Z4}
\end{table}

Now, we derive the $\mathbb{Z}_4$-twisted basis explicitly from three $\mathbb{Z}_2$-even modes of $M=5$.
In this case, one of eigenvalues is $p=-1$ and the other two eigenvalues are $p=1$.
The $\mathbb{Z}_4$-twisted eigenstates are obtained by
\begin{align}
\psi_{T^2/\mathbb{Z}_4^p} &= {\cal N} \left(
\begin{pmatrix}
1 & & \\
 & 1 & \\
 & & 1
\end{pmatrix}
+ \frac{ p^{-1}}{\sqrt{5}}
\begin{pmatrix}
1 & \sqrt{2} & \sqrt{2} \\
\sqrt{2} & \frac{\sqrt{5}-1}{2} & -\frac{\sqrt{5}+1}{2} \\
\sqrt{2} & -\frac{\sqrt{5}+1}{2} & \frac{\sqrt{5}-1}{2}
\end{pmatrix}
\right)
\begin{pmatrix}
\psi^{(0,0),5}_{T^2/\mathbb{Z}_2^{+}} \\
\psi^{(1,0),5}_{T^2/\mathbb{Z}_2^{+}} \\
\psi^{(2,0),5}_{T^2/\mathbb{Z}_2^{+}}
\end{pmatrix},
\end{align}
where
\begin{align}
\begin{aligned}
\psi^{(0,0),5}_{T^2/\mathbb{Z}_2^{+}} &= \psi^{(0,0),5}_{T^2}, \\
\psi^{(1,0),5}_{T^2/\mathbb{Z}_2^{+}} &= \frac{1}{\sqrt{2}} (\psi^{(1,0),5}_{T^2} + \psi^{(4,0),5}_{T^2}), \\
\psi^{(2,0),5}_{T^2/\mathbb{Z}_2^{+}} &= \frac{1}{\sqrt{2}} (\psi^{(2,0),5}_{T^2} + \psi^{(3,0),5}_{T^2}).
\end{aligned}
\end{align}
Then, the mode with $p=-1$ is obtained as
\begin{align}
\psi^1_{T^2/\mathbb{Z}_4^{-1}}
&= {\cal N}' \left( \frac{\sqrt{5}-1}{\sqrt{2}} \psi^{(0,0),5}_{T^2/\mathbb{Z}_2^{+}} - \psi^{(1,0),5}_{T^2/\mathbb{Z}_2^{+}} - \psi^{(2,0),5}_{T^2/\mathbb{Z}_2^{+}} \right) \notag \\
&= \sqrt{\frac{3-\sqrt{5}}{5-\sqrt{5}}} \psi^{(0,0),5}_{T^2/\mathbb{Z}_2^{+}} - \frac{1}{\sqrt{5-\sqrt{5}}} \psi^{(1,0),5}_{T^2/\mathbb{Z}_2^{+}} - \frac{1}{\sqrt{5-\sqrt{5}}} \psi^{(2,0),5}_{T^2/\mathbb{Z}_2^{+}}, \quad ({\cal N}'=\frac{1}{\sqrt{5-\sqrt{5}}}), \label{p-1}
\end{align}
while one mode with $p=1$ is obtained as
\begin{align}
\psi^1_{T^2/\mathbb{Z}_4^{+1}}
&= {\cal N}' \left( \sqrt{2} \psi^{(0,0),5}_{T^2/\mathbb{Z}_2^{+}} + \frac{3\sqrt{5}-1}{2} \psi^{(1,0),5}_{T^2/\mathbb{Z}_2^{+}} - \frac{\sqrt{5}+1}{2} \psi^{(2,0),5}_{T^2/\mathbb{Z}_2^{+}} \right) \notag \\
&= \frac{2}{\sqrt{30-2\sqrt{5}}} \psi^{(0,0),5}_{T^2/\mathbb{Z}_2^{+}} + \sqrt{\frac{23-3\sqrt{5}}{30-2\sqrt{5}}} \psi^{(1,0),5}_{T^2/\mathbb{Z}_2^{+}} - \sqrt{\frac{3+\sqrt{5}}{30-2\sqrt{5}}} \psi^{(2,0),5}_{T^2/\mathbb{Z}_2^{+}}, \quad ({\cal N}'=\frac{1}{\sqrt{15-\sqrt{5}}}). \label{p+1}
\end{align}
The other mode with $p=1$ is obtained by orthogonalizing
\begin{align}
\psi_{T^2/\mathbb{Z}_4^{+1}}
&= {\cal N}' \left( \frac{\sqrt{5}+1}{\sqrt{2}} \psi^{(0,0),5}_{T^2/\mathbb{Z}_2^{+}} + \psi^{(1,0),5}_{T^2/\mathbb{Z}_2^{+}} + \psi^{(2,0),5}_{T^2/\mathbb{Z}_2^{+}} \right) \notag
\end{align}
to one in Eq.~(\ref{p+1}) through the Gram-Schmidt process as the following,
\begin{align}
\psi^2_{T^2/\mathbb{Z}_4^{+1}}
&= \sqrt{\frac{3+\sqrt{5}}{7+\sqrt{5}}} \psi^{(0,0),5}_{T^2/\mathbb{Z}_2^{+}} + \frac{2}{\sqrt{7+\sqrt{5}}} \psi^{(2,0),5}_{T^2/\mathbb{Z}_2^{+}}. \label{p+2}
\end{align}
Therefore, the $\mathbb{Z}_4$-twisted basis obtained from the $\mathbb{Z}_2$-twisted even modes of $M=5$, $\psi_{T^2/\mathbb{Z}_4} = P^{\mathrm T} \psi_{T^2/\mathbb{Z}_2}$, are given as
\begin{align}
\begin{pmatrix}
\psi^1_{T^2/\mathbb{Z}_4^{+1}} \\ \psi^1_{T^2/\mathbb{Z}_4^{-1}} \\ \psi^2_{T^2/\mathbb{Z}_4^{+1}}
\end{pmatrix}
=
\begin{pmatrix}
\sqrt{\frac{23-3\sqrt{5}}{30-2\sqrt{5}}} & \frac{2}{\sqrt{30-2\sqrt{5}}} & - \sqrt{\frac{3+\sqrt{5}}{30-2\sqrt{5}}} \\
- \frac{1}{\sqrt{5-\sqrt{5}}} & \sqrt{\frac{3-\sqrt{5}}{5-\sqrt{5}}} & - \frac{1}{\sqrt{5-\sqrt{5}}} \\
0 & \sqrt{\frac{3+\sqrt{5}}{7+\sqrt{5}}} & \frac{2}{\sqrt{7+\sqrt{5}}} \\
\end{pmatrix}
\begin{pmatrix}
\psi^{(1,0),5}_{T^2/\mathbb{Z}_2^{+}} \\ \psi^{(0,0),5}_{T^2/\mathbb{Z}_2^{+}} \\ \psi^{(2,0),5}_{T^2/\mathbb{Z}_2^{+}}
\end{pmatrix}.
\end{align}
This gives $P$ as Eq.~(\ref{Z4Z2}).



\begin{thebibliography}{99}



\bibitem{Bachas:1995ik}
  C.~Bachas,
  arXiv:hep-th/9503030.

\bibitem{Blumenhagen:2000wh}
  R.~Blumenhagen, L.~Goerlich, B.~Kors and D.~Lust,
  JHEP {\bf 0010}, 006 (2000)
  [arXiv:hep-th/0007024].

\bibitem{Angelantonj:2000hi}
  C.~Angelantonj, I.~Antoniadis, E.~Dudas and A.~Sagnotti,
  Phys. Lett. {\bf B489}, 223 (2000)
  [arXiv:hep-th/0007090].

\bibitem{Blumenhagen:2000ea}
  R.~Blumenhagen, B.~Kors and D.~Lust,
  JHEP {\bf 0102}, 030 (2001)
  [arXiv:hep-th/0012156].



\bibitem{Cremades:2004wa}
D.~Cremades, L.~E.~Ibanez and F.~Marchesano,
JHEP \textbf{05} (2004), 079
[arXiv:hep-th/0404229 [hep-th]].


\bibitem{Abe:2009dr}
H.~Abe, K.~S.~Choi, T.~Kobayashi and H.~Ohki,
JHEP \textbf{06}, 080 (2009)
[arXiv:0903.3800 [hep-th]].




\bibitem{Abe:2008fi}
H.~Abe, T.~Kobayashi and H.~Ohki,
JHEP \textbf{09} (2008), 043
[arXiv:0806.4748 [hep-th]].


\bibitem{Abe:2013bca} 
  T.~H.~Abe, Y.~Fujimoto, T.~Kobayashi, T.~Miura, K.~Nishiwaki and M.~Sakamoto,
  JHEP {\bf 1401}, 065 (2014)
  [arXiv:1309.4925 [hep-th]].
 
  
\bibitem{Abe:2014noa} 
  T.~h.~Abe, Y.~Fujimoto, T.~Kobayashi, T.~Miura, K.~Nishiwaki and M.~Sakamoto,
  Nucl.\ Phys.\ B {\bf 890}, 442 (2014)
  [arXiv:1409.5421 [hep-th]].
 




\bibitem{Kobayashi:2017dyu} 
 T.~Kobayashi and S.~Nagamoto,
 Phys.\ Rev.\ D {\bf 96}, no. 9, 096011 (2017)
 [arXiv:1709.09784 [hep-th]].


\bibitem{Sakamoto:2020pev}
M.~Sakamoto, M.~Takeuchi and Y.~Tatsuta,
Phys. Rev. D \textbf{102}, no.2, 025008 (2020)
[arXiv:2004.05570 [hep-th]]; 
[arXiv:2010.14214 [hep-th]].




\bibitem{Fujimoto:2013xha}
Y.~Fujimoto, T.~Kobayashi, T.~Miura, K.~Nishiwaki and M.~Sakamoto,
Phys.\ Rev.\ D \textbf{87} (2013) no.8, 086001
[arXiv:1302.5768 [hep-th]].









\bibitem{Abe:2008sx}
H.~Abe, K.~S.~Choi, T.~Kobayashi and H.~Ohki,
Nucl. Phys. B \textbf{814}, 265-292 (2009)
[arXiv:0812.3534 [hep-th]].


\bibitem{Abe:2015yva}
T.~h.~Abe, Y.~Fujimoto, T.~Kobayashi, T.~Miura, K.~Nishiwaki, M.~Sakamoto and Y.~Tatsuta,
Nucl. Phys. B \textbf{894}, 374-406 (2015)
[arXiv:1501.02787 [hep-ph]].



\bibitem{Hoshiya:2020hki}
K.~Hoshiya, S.~Kikuchi, T.~Kobayashi, Y.~Ogawa and H.~Uchida,
[arXiv:2012.00751 [hep-th]].






\bibitem{Abe:2012fj}
H.~Abe, T.~Kobayashi, H.~Ohki, A.~Oikawa and K.~Sumita,
Nucl. Phys. B \textbf{870}, 30-54 (2013)
[arXiv:1211.4317 [hep-ph]].  

\bibitem{Abe:2014vza}
H.~Abe, T.~Kobayashi, K.~Sumita and Y.~Tatsuta,
Phys. Rev. D \textbf{90}, no.10, 105006 (2014)
[arXiv:1405.5012 [hep-ph]].





\bibitem{Fujimoto:2016zjs}
Y.~Fujimoto, T.~Kobayashi, K.~Nishiwaki, M.~Sakamoto and Y.~Tatsuta,
Phys. Rev. D \textbf{94}, no.3, 035031 (2016)
[arXiv:1605.00140 [hep-ph]].




\bibitem{Kobayashi:2016qag}
T.~Kobayashi, K.~Nishiwaki and Y.~Tatsuta,
JHEP \textbf{04}, 080 (2017)
[arXiv:1609.08608 [hep-th]].









\bibitem{Kobayashi:2018rad} 
 T.~Kobayashi, S.~Nagamoto, S.~Takada, S.~Tamba and T.~H.~Tatsuishi,
 Phys.\ Rev.\ D {\bf 97}, no. 11, 116002 (2018)
 [arXiv:1804.06644 [hep-th]].

\bibitem{Kobayashi:2018bff}
T.~Kobayashi and S.~Tamba,
Phys.\ Rev.\ D {\bf 99} (2019) no.4, 046001
[arXiv:1811.11384 [hep-th]].

\bibitem{Kariyazono:2019ehj}
Y.~Kariyazono, T.~Kobayashi, S.~Takada, S.~Tamba and H.~Uchida,
Phys. Rev. D \textbf{100}, no.4, 045014 (2019)
[arXiv:1904.07546 [hep-th]].

\bibitem{Ohki:2020bpo}
H.~Ohki, S.~Uemura and R.~Watanabe,
Phys. Rev. D \textbf{102}, no.8, 085008 (2020)
[arXiv:2003.04174 [hep-th]].



\bibitem{Kikuchi:2020frp}
S.~Kikuchi, T.~Kobayashi, S.~Takada, T.~H.~Tatsuishi and H.~Uchida,
Phys. Rev. D \textbf{102}, no.10, 105010 (2020)
[arXiv:2005.12642 [hep-th]].


\bibitem{Kikuchi:2020nxn}
S.~Kikuchi, T.~Kobayashi, H.~Otsuka, S.~Takada and H.~Uchida,
JHEP \textbf{11}, 101 (2020)
[arXiv:2007.06188 [hep-th]].





\bibitem{Kikuchi:2021ogn}
S.~Kikuchi, T.~Kobayashi and H.~Uchida,
[arXiv:2101.00826 [hep-th]].

\bibitem{Almumin:2021fbk}
Y.~Almumin, M.~C.~Chen, V.~Knapp-Perez, S.~Ramos-Sanchez, M.~Ratz and S.~Shukla,
[arXiv:2102.11286 [hep-th]].



\bibitem{Blumenhagen:2006xt}
R.~Blumenhagen, M.~Cvetic and T.~Weigand,
Nucl. Phys. B \textbf{771}, 113-142 (2007)
[arXiv:hep-th/0609191 [hep-th]].


\bibitem{Ibanez:2006da}
L.~E.~Ibanez and A.~M.~Uranga,
JHEP \textbf{03}, 052 (2007)
[arXiv:hep-th/0609213 [hep-th]].

\bibitem{Ibanez:2007rs}
L.~E.~Ibanez, A.~N.~Schellekens and A.~M.~Uranga,
JHEP \textbf{06}, 011 (2007)
[arXiv:0704.1079 [hep-th]].

\bibitem{Antusch:2007jd}
S.~Antusch, L.~E.~Ibanez and T.~Macri,
JHEP \textbf{09}, 087 (2007)
[arXiv:0706.2132 [hep-ph]].

\bibitem{Kobayashi:2015siy}
T.~Kobayashi, Y.~Tatsuta and S.~Uemura,
Phys. Rev. D \textbf{93}, no.6, 065029 (2016)
[arXiv:1511.09256 [hep-ph]].



\bibitem{Ishiguro:2020tmo}
K.~Ishiguro, T.~Kobayashi and H.~Otsuka,
[arXiv:2011.09154 [hep-ph]].



\bibitem{Feruglio:2017spp} 
  F.~Feruglio,
  arXiv:1706.08749 [hep-ph];
%
  T.~Kobayashi, K.~Tanaka and T.~H.~Tatsuishi,
  Phys.\ Rev.\ D {\bf 98}, no. 1, 016004 (2018)
  [arXiv:1803.10391 [hep-ph]];
%
  J.~T.~Penedo and S.~T.~Petcov,
  Nucl.\ Phys.\ B {\bf 939}, 292 (2019)
  [arXiv:1806.11040 [hep-ph]];
%
  J.~C.~Criado and F.~Feruglio,
  SciPost Phys.\  {\bf 5}, no. 5, 042 (2018)
  [arXiv:1807.01125 [hep-ph]];
%
  T.~Kobayashi, N.~Omoto, Y.~Shimizu, K.~Takagi, M.~Tanimoto and T.~H.~Tatsuishi,
  JHEP {\bf 1811}, 196 (2018)
  [arXiv:1808.03012 [hep-ph]];
%
  P.~P.~Novichkov, J.~T.~Penedo, S.~T.~Petcov and A.~V.~Titov,
  JHEP {\bf 1904}, 005 (2019)
  [arXiv:1811.04933 [hep-ph]];
%
  JHEP {\bf 1904}, 174 (2019)
  [arXiv:1812.02158 [hep-ph]];
%
  F.~J.~de Anda, S.~F.~King and E.~Perdomo,
  arXiv:1812.05620 [hep-ph];
%
  H.~Okada and M.~Tanimoto,
  Phys.\ Lett.\ B {\bf 791}, 54 (2019)
  [arXiv:1812.09677 [hep-ph]];
%
  T.~Kobayashi, Y.~Shimizu, K.~Takagi, M.~Tanimoto, T.~H.~Tatsuishi and H.~Uchida,
  Phys.\ Lett.\ B {\bf 794}, 114 (2019)
  [arXiv:1812.11072 [hep-ph]];
%
  P.~P.~Novichkov, S.~T.~Petcov and M.~Tanimoto,
  Phys.\ Lett.\ B {\bf 793}, 247 (2019)
  [arXiv:1812.11289 [hep-ph]].





\bibitem{Kobayashi:2019fma}
T.~Kobayashi, H.~Otsuka and H.~Uchida,
JHEP \textbf{08}, 046 (2019)
[arXiv:1904.02867 [hep-th]];
%
JHEP \textbf{03}, 042 (2020)
[arXiv:1911.01930 [hep-ph]].



	\bibitem{Gunning:1962}
 	R. C. Gunning,
 	\textit{Lectures on Modular Forms}
 	(Princeton University Press, Princeton, NJ, 1962).

\bibitem{Schoeneberg:1974}
 B.~Schoeneberg,
  \textit{Elliptic Modular Functions}
  (Springer-Verlag, 1974)

\bibitem{Koblitz:1984}
 N.~Koblitz,
  \textit{Introduction to Elliptic Curves and Modular Forms}
  (Springer-Verlag, 1984)

\bibitem{Bruinier:2008}
J.H.~Bruinier, G.V.D.~Geer, G.~Harder, and D.~Zagier,
\textit{The 1-2-3 of Modular Forms}
(Springer, 2008)

\bibitem{Baur:2019kwi} 
 A.~Baur, H.~P.~Nilles, A.~Trautner and P.~K.~S.~Vaudrevange,
 Phys.\ Lett.\ B {\bf 795}, 7 (2019)
 [arXiv:1901.03251 [hep-th]].


\bibitem{Novichkov:2019sqv} 
  P.~P.~Novichkov, J.~T.~Penedo, S.~T.~Petcov and A.~V.~Titov,
  JHEP {\bf 1907}, 165 (2019)
  [arXiv:1905.11970 [hep-ph]].

\bibitem{deAdelhartToorop:2011re} 
R.~de Adelhart Toorop, F.~Feruglio and C.~Hagedorn,
Nucl.\ Phys.\ B {\bf 858}, 437 (2012)
[arXiv:1112.1340 [hep-ph]].

\bibitem{Liu:2019khw}
X.~G.~Liu and G.~J.~Ding,
JHEP {\bf 1908}, 134 (2019)
[arXiv:1907.01488 [hep-ph]];



\end{thebibliography}
\end{document}